\documentclass[onecolumn,amsmath,amssymb,subfigure,12pt]{revtex4-1}
\usepackage{bm}
\usepackage{graphicx}
\usepackage{hhline}
\usepackage[newitem,newenum]{paralist}
 {\pointedenum\begin{enumerate}}%
 {\end{enumerate}}
 {\pointlessenum\begin{enumerate}}%
 {\end{enumerate}} 
\usepackage{mathrsfs}
\usepackage{amssymb}
\usepackage{bm}
\usepackage[dvips]{color}
\usepackage{color}
\usepackage{fancybox}
\usepackage{fancybox}
\usepackage{fancyhdr}
\begin{document}
\title{From conduction to convection of thermally relativistic fluids between two parallel walls under gravitational force}
\author{Ryosuke Yano}
\affiliation{Department of Advanced Energy, University of Tokyo, 5-1-5 Kashiwanoha, Kashiwa, Chiba 277-8561, Japan}
\email{yano@k.u-tokyo.ac.jp}
\begin{abstract}
We discuss the thermal conduction and convection of thermally relativistic fluids between two parallel walls under the gravitational force, both theoretically and numerically. In the theoretical discussion, we assume that the Lorentz contraction is ignored and spacetime is flat. For understanding of the thermal conduction and convection of thermally relativistic fluids between two parallel walls under the gravitational force, we solve the relativistic Boltzmann equation using the direct simulation Monte Carlo method. Numerical results indicate that strongly nonequilibrium states are formed in vicinities of two walls, which do not allow us to discuss \textcolor{black}{the transition of the thermal conduction to the thermal convection} of thermally relativistic fluids under the gravitational force \textcolor{black}{in the framework of the relativistic Navier-Stokes-Fourier equation}, when the flow-field is under the transition regime between the rarefied and continuum regimes, whereas such strongly nonequilibrium states are not formed in vicinities of two walls under the nonrelativistic limit. 
\end{abstract}
\maketitle
\section{Introduction}
In the early epoch of universe, the space is filled with high energy partons, whose thermal energy is larger than $\Lambda_{\mbox{\tiny{QCD}}}$ \cite{Peskin}, which enables us to assume the asymptotic freedom of partons. In this paper, we assume that partons are approximately expressed with hard spheres under the asymptotic freedom of partons in the early epoch of universe to simplify our discussion. Of course, the collisional differential cross section depends on the momentum transferred between two colliding partons, and the collisional deflection-angle also depends on the momentum, which is transferred between two colliding partons in the energy-regime of the asymptotic freedom \cite{Peskin} \cite{Hatsuda}. In particular, we characterize the thermal energy of partons using the thermally relativistic measure, namely, $\chi=mc^2/\left(k\theta\right)$ ($m$: mass, $c$: speed of light, $\theta$: temperature, $k$: Boltzmann constant), whereas we assume that $m$ is large enough to realize $\chi \rightarrow 0$ by not $m \rightarrow 0$ but $\theta \rightarrow \infty$. Additionally, $\Lambda_{\mbox{\tiny{QCD}}} \ll mc^2$ is assumed. Consequently, thermally relativistic fluids are characterized by $\chi \le 100$. Additionally, we assume that the asymptotic freedom of partons always consists under $\chi \le 100$.\\ 
In our previous studies, we discussed thermally relativistic fluids with dissipation, which are composed of hard spherical partons, such as the shock layer or thermal fluctuation under the equilibrium state \cite{Yano}. Meanwhile, the characteristics of the thermal conduction and convection of thermally relativistic fluids under the gravitational force, which are composed of hard spherical partons, have not been understood, adequately. In this paper, we investigate the thermal conduction and convection of thermally relativistic fluids under the gravitational force in flat spacetime for understanding of the thermal conduction and convection of thermally relativistic fluids under the gravitational force in the early epoch of the universe. Of course, such an assumption of flat spacetime is inadequate to discuss the thermal conduction and convection of thermally relativistic fluids under the gravitational force in the early epoch of universe, because spacetime is markedly curved in the early epoch of universe \cite{Yano2}. We, however, consider that our present study of the thermal conduction and convection of thermally relativistic fluids under the gravitational force in flat spacetime will be useful for understanding of the thermal conduction and convection of thermally relativistic fluids under curved spacetime in the early epoch of universe.\\
The thermal conduction and convection of thermally relativistic fluids under the gravitational force in flat spacetime are studied by investigating the thermal conduction and convection between two parallel walls, which exist at $x=0$ and $L$, as shown in Fig. 1. Temperatures of two walls ($\theta_{-}^w$: temperature of the cold wall at $x_1=0$, $\theta_{+}^w$: temperature of the hot wall at $x_1=L$) are different, namely, $\theta_{-}^w<\theta_{+}^w$, whereas the gravitational force is defined by $\bm{g}=\left(g_1,g_2,g_3\right)=\left(g,0,0\right)$ $\left(0<g\right)$. In particular, the thermal convection via such a gravitational force is called as Rayleigh B$\grave{\mbox{e}}$nard convection \cite{Stefanov}. For example, two parallel walls correspond to thermal bathes (energy sources) in the early epoch of universe.\\
In this paper, we investigate the characteristics of the thermal conduction and thermal convection, namely, Rayleigh B$\grave{\mbox{e}}$nard convection of thermally relativistic fluids, both theoretically and numerically, whereas effects of Lorentz contraction are ignorable owing to $\left|\bm{u}\right|\ll c$ ($\bm{u}$: flow velocity). Our previous studies \cite{Yano} indicated that the relativistic Navier-Stokes-Fourier (NSF) equation is efficient for the analytical comprehension of the weakly nonequilibrium state such as thermally fluctuations under the static state, when the Lorentz contraction is ignorable and thermally relativistic effects are dominant.\\ 
This paper is organized as follows. In Sec. II, we define the Rayleigh number for thermally relativistic fluids on the basis of the relativistic NSF equation to introduce the theoretical value of the critical Rayleigh number in appendix A. \textcolor{black}{In Sec. II, temperatures on two walls, which define the Rayleigh number, are not defined, analytically.} \textcolor{black}{Thus, we consider temperature jumps on two walls on the basis of Grad's 14 moments equation coupled with Fourier law in Sec. III. Here, we must notice that temperature jumps on two walls are calculated using the method \cite{Struchtrup}, which has been used to calculate the temperature jump of the non-relativistic gas on the wall as a half-space problem \cite{Cercignani2}, whereas Israel-Stewart equation, which was applied to describe the boundless Truesdell's uniform shear flow by Yano-Suzuki \cite{Yano3}, cannot be applied to describe the profiles of Grad's 14 moments on the wall, because the distribution function is discontinuous on the wall.} In Sec. III-A, we find analytical solutions of the density and temperature under the pure conductive state under two limits, namely, nonrelativistic limit ($\chi \rightarrow \infty$) and thermally relativistic limit ($\chi \rightarrow 0$). \textcolor{black}{Finally, we calculate temperature jumps on two walls in the framework of Grad's 14 moments coupled with the Fourier law under two limits using the analytical solution of the temperature under the pure conductive state.} In Sec. IV, we numerically investigate the characteristics of the thermal conduction and convection of thermally relativistic fluids under the gravitational force by solving the relativistic Boltzmann equation on the basis of direct simulation Monte Carlo (DSMC) method \cite {Bird} \cite{Yano}. Numerical results, however, indicate that \textcolor{black}{the transition of the thermal conduction to the thermal convection} of thermally relativistic fluids under the gravitational force cannot be discussed in the framework of \textcolor{black}{the relativistic NSF equation} owing to strongly nonequilibrium states in vicinities of cold and hot walls, when the flow-field is under the transition regime between the rarefied and continuum regimes. \textcolor{black}{In particular, the temperature jumps on two walls increase in accordance with the decrease in the temperature, because the representative relaxation rate \cite{Cercignani} decreases in accordance with the decrease in the temperature, when Knudsen number is fixed}. As a result, we cannot use the Rayleigh number as a parameter, which characterizes the transition of the thermal conduction to the thermal convection, \textcolor{black}{under the transition regime between the rarefied and continuum regimes}. Finally, we make concluding remarks in Sec. V.
\section{Definition of Rayleigh number for thermally relativistic fluids}
The relativistic NSF equation is written for thermally relativistic fluids with zero Lorentz contraction in acausal form as \cite{Yano}
\begin{eqnarray}
\frac{\partial \rho}{\partial t}&=&-\bm{\nabla} \cdot \left(\rho \bm{u}\right),\\
\frac{\partial \rho u_i G}{\partial t}&=&-\frac{\partial}{\partial x_k}\left(p \delta_{ik}+\rho u_i u_k G+\Pi_{ik}\right)+\rho g_i,\\
\rho c_v \frac{\partial \theta}{\partial t}&=&-\rho c_v \bm{u}\cdot\bm{\nabla} \theta-\bm{\nabla} \cdot \bm{q}-p \bm{\nabla} \cdot \bm{u},
\end{eqnarray}
where $\rho$ is the density, $\bm{u}=\left(u_1,u_2,u_3\right)$ is the flow velocity, $p$ is the static pressure, $g_i$ is the gravitational components, $G\equiv K_3\left(\chi\right)/K_2\left(\chi\right)$, in which $K_n$ is the n-th order modified Bessel function of the second kind, $\Pi_{ij}$ is the deviatoric stress tensor, $\bm{q}=\left(q_1,q_2,q_3\right)$ is the heat flux vector, and $c_v=k\left(\chi^2+5G\chi-G^2\chi^2-1\right)$ is the specific heat at the constant volume. The derivation of Eqs. (1)-(3) is described in appendix C in detail. The formal difference between the thermally relativistic NSF equation and nonrelativistic NSF equation is a term $G$ in Eq. (2). Consequently, the form of thermally relativistic NSF equation coincides with the form of the nonrelativistic NSF equation, when $G=1$.\\
$\Pi_{ij}$ \textcolor{black}{in Eq. (2)} and $\bm{q}$ \textcolor{black}{in Eq. (3)} are written in the framework of the NSF law as follows \cite{Yano}
\begin{eqnarray}
\Pi_{ij}&=&-\eta\left(\frac{\partial u_i}{\partial x_j}+\frac{\partial u_j}{\partial x_i}\right)-\left(\zeta-\frac{2}{3}\eta\right)\delta_{ij}\frac{\partial u_l}{\partial x_l},\\
\bm{q}&=&-\lambda \bm{\nabla}\theta,
\end{eqnarray}
where $\eta$ is the viscosity coefficient, $\zeta$ is the bulk viscosity, and $\lambda$ is the thermal conductivity, which are calculated for hard spherical partons by Cercignani and Kremer \cite{Cercignani} on the basis of Israel-Stewart theory.\\
Provided that the flow is incompressible, we can rewrite Eqs. (1)-(3) as
\begin{eqnarray}
\bm{\nabla} \cdot \bm{u}&=&0,\\
\rho \frac{\partial u_i G}{\partial t}&=&-\bm{\nabla} p -\rho u_k \frac{\partial}{\partial x_k} u_i G+\eta \nabla^2 u_i+\rho g_i,\\
\rho c_v \frac{\partial \theta}{\partial t}&=&-\rho c_v \bm{u}\cdot\bm{\nabla} \theta+\lambda \nabla^2 \theta-p \bm{\nabla} \cdot \bm{u},
\end{eqnarray}
Next, we assume that quantities can be expressed with those under the background state and their perturbations as
\begin{eqnarray}
\rho&=&\rho_0+\delta \rho,~~~\bm{u}=\bm{u}_0+\delta \bm{u},~~~p=p_0+\delta p,\nonumber\\
\theta&=&\theta_0+\delta \theta,~~~G=G_0+\delta G\simeq G_0-G_0\frac{\delta \theta}{\theta_0},
\end{eqnarray}
where $\theta_0=\theta_{-}$ and $p_0=\rho_0 R \theta_0$ ($R=k/m$), in which $\theta_{-}=\theta|_{x_1=0}$.\\
Equation (7) yields the following relation under the static state, namely, $\bm{u}=0$
\begin{eqnarray}
\rho_0 \bm{g}=\bm{\nabla} p_0,
\end{eqnarray}
Subtracting Eq. (10) from Eq. (7) and dividing both sides of Eq. (7) by $\rho_0G_0$, we rewrite Eq. (7) using Eq. (9) as
\begin{eqnarray}
&&\left(1+\delta \tilde{\rho} \right)\frac{\partial \delta u_i}{\partial t} \nonumber \\
&&=-\bm{\nabla} \delta \tilde{p}-\left(1+\delta \tilde{\rho} \right)\delta u_k\frac{\partial}{\partial x_k} \delta u_i+\tilde{\eta} \nabla^2 \delta u_i+\delta \tilde{\rho} \tilde{g}_i
\end{eqnarray}
where we used approximations $\delta u_i \left(\frac{\partial \delta G}{\partial t}+u_k \frac{\partial \delta G}{\partial x_k}\right)\sim 0$ and $\delta \rho \delta G g_i \sim 0$, and relations $\delta \tilde{\rho}=\delta \rho/\rho_0$, $\delta \tilde{G}=\delta G/G_0$, $\delta \tilde{p}=\delta p/\left(\rho_0 G_0\right)$, $\tilde{\eta}=\eta/\left(\rho_0 G_0\right)$, and $\tilde{g}_i=g_i/G_0$.\\
Setting $\delta \tilde{\rho} \sim 0$ in the first term of the left hand side of Eq. (11) and second term of the right hand side of Eq. (11), Eq. (11) is rewritten as
\begin{eqnarray}
&&\frac{\partial \delta u_i}{\partial t}=-\bm{\nabla} \delta \tilde{p}-\delta u_k\frac{\partial}{\partial x_k} \delta u_i+\tilde{\eta} \nabla^2 \delta u_i+\delta \tilde{\rho} \tilde{g}_i,
\end{eqnarray}
Substituting Eq. (9) into Eq. (8), we obtain the linear balance equation of energy by neglecting nonlinear terms
\begin{eqnarray}
\frac{\partial \delta \theta}{\partial t}=\tilde{\lambda} \nabla^2 \delta \theta-\delta \bm{u} \cdot \nabla \delta \theta,
\end{eqnarray}
where we used the relation $-p \bm{\nabla} \cdot \bm{v}=-\rho \left(c_p-c_v\right)\left(\frac{\partial \theta}{\partial t}+\bm{u} \cdot \nabla \theta\right)$ and $\tilde{\lambda}=\lambda/\left(c_p \rho_0\right)$.\\
To nondimensionalize Eqs. (12) and (13), we introduce following relations
\begin{eqnarray}
\delta \tilde{p}=\mathscr{U}^2 \frac{\tilde{\eta}}{\tilde{\lambda}} \delta \bar{p},~~~\delta \bm{u}=\mathscr{U} \delta \bar{\bm{u}},~~~\delta \theta=\left(\theta_1-\theta_0\right) \delta \bar{\theta},~~~\bm{x}=L \bar{\bm{x}},~~~t=\frac{L}{\mathscr{U}} \bar{t},
\end{eqnarray}
where $\theta_1=\theta_+$, in which $\theta_+=\theta|_{x_1=L}$.\\
Substituting Eq. (14) into Eq. (12), we obtain using $\bar{\bm{\nabla}}=L \bm{\nabla}$ as
\begin{eqnarray}
\frac{\mathscr{U}^2}{L}\frac{\partial \delta \bar{u}_i}{\partial \bar{t}}=-\frac{\mathscr{U}^2}{L}\delta \bar{u}_k\frac{\partial}{\partial \bar{x}_k} \delta \bar{u}_i-\frac{\mathscr{U}^2}{L} \frac{\Pr}{G_0} \bar{\bm{\nabla}} \delta \bar{p}+\frac{\mathscr{U}^2}{L} \tilde{\eta} \bar{\nabla}^2 \delta \bar{u}_i+\delta \tilde{\rho} \tilde{g}_i,
\end{eqnarray}
\textcolor{black}{where $\Pr=\eta c_p/\lambda$ is Prandtl number}.\\
Here, we use Boussinesq's approximation:
\begin{eqnarray}
\rho=\rho_0\left\{1+\alpha\left(\theta_1-\theta_0\right)\delta \bar{\theta}\right\},
\end{eqnarray}
where $\alpha=-1/\rho_0 (\partial \rho/\partial \theta)_{p=p_0}$.\\
From Eq. (16), we readily obtain
\begin{eqnarray}
\delta \tilde{\rho}=\frac{\rho-\rho_0}{\rho_0}=\alpha\left(\theta_1-\theta_0\right)\delta \bar{\theta}.
\end{eqnarray}
Substituting Eq. (17) into Eq. (15) and dividing both sides of Eq. (15) by $\mathscr{U}^2/L$, we obtain
\begin{eqnarray}
\frac{\partial \delta \bar{u}_i}{\partial \bar{t}}=-\delta \bar{u}_k\frac{\partial}{\partial \bar{x}_k} \delta \bar{u}_i-\frac{\Pr}{G_0} \bar{\bm{\nabla}} \delta \bar{p}+\frac{\tilde{\eta}}{\mathscr{U} L} \bar{\nabla}^2 \delta \bar{u}_i+\frac{L}{\mathscr{U}^2}\alpha\left(\theta_1-\theta_0\right)\delta \bar{\theta} \tilde{g}_i,
\end{eqnarray}
\textcolor{black}{Here, we rewrite Eq. (18) using the relation $\mathscr{U}=\tilde{\lambda}/L$ as}
\begin{eqnarray}
\frac{\partial \delta \bar{u}_i}{\partial \bar{t}}&=&-\delta \bar{u}_k\frac{\partial}{\partial \bar{x}_k} \delta \bar{u}_i-\hat{\Pr} \bar{\bm{\nabla}} \delta \bar{p}^\prime+\hat{\Pr}\bar{\nabla}^2 \delta \bar{u}_i+\hat{\Pr} \frac{L^3}{\tilde{\lambda}\tilde{\eta}}\alpha\left(\theta_1-\theta_0\right)\delta \bar{\theta} \tilde{g}_i,
\end{eqnarray}
where $\delta p^\prime=\delta p/\rho_0$ and $\hat{\Pr}=\Pr/G_0$.\\
Dividing both sides of Eq. (19) by $\hat{\mbox{Pr}}$, we obtain
\begin{eqnarray}
\frac{1}{\hat{\Pr}}\frac{\partial \delta \bar{u}_i}{\partial \bar{t}}=-\frac{1}{\hat{\Pr}}\delta \bar{u}_k\frac{\partial}{\partial \bar{x}_k} \delta \bar{u}_i-\bar{\bm{\nabla}} \delta \bar{p}^\prime+\bar{\nabla}^2 \delta \bar{u}_i+\mbox{Ra}\delta \bar{\theta} e_i,
\end{eqnarray}
where $\mbox{Ra}=\tilde{g}L^3\alpha\left(\theta_1-\theta_0\right)/\left(\tilde{\lambda}\tilde{\eta}\right)$ is the Rayleigh number for thermally relativistic fluids, and $\bm{e}$ is the unit vector.\\
Finally, Ra for thermally relativistic fluids is
\begin{eqnarray}
\mbox{Ra}=\frac{g L^3\rho_0^2 c_p \alpha\left(\theta_1-\theta_0\right)}{\eta \lambda},
\end{eqnarray}
where $c_p=k\left(\chi^2+5G\chi-G^2 \chi^2\right)$ is the specific heat at constant pressure.\\
Provided that thermally relativistic fluids are composed of hard spherical partons with mass $m$ and diameter $d$, we obtain Ra using definitions of $\eta$ and $\lambda$ for hard spherical partons, which were calculated by Cercignani and Kremer on the basis of Israel-Stewart theory \cite{Cercignani}, as
\begin{eqnarray}
\mbox{Ra}&=&\frac{256}{45} \frac{\psi}{\mbox{Fr} \mbox{Kn}^2} \left(1-r\right), ~~~\left(r=\theta_0/\theta_1\right)\nonumber \\
\psi&=&\frac{\left\{\left(\chi^2+2\right) K_2(2\chi)+5 \chi K_3(2 \chi)\right\} \left\{\left(15 \chi^2+2\right) K_2(2 \chi)+\chi \left(3 \chi^2+49\right) K_3(2 \chi)\right\}}{\chi^7 K_3(\chi){}^2 \left(\chi K_2(\chi){}^2+5 K_3(\chi) K_2(\chi)-\chi K_3(\chi){}^2\right)} \nonumber \\
&=& 0.29~~ \left(\chi \rightarrow 0\right) \nonumber \\
&& 0.38~~ \left(\chi \rightarrow \infty \right),
\end{eqnarray}
where $\mbox{Fr}=\left(2k\theta_1/m\right)/\left(gL\right)$ is Froude number, $\mbox{Kn}=1/\left(\sqrt{2} n_0\pi d^2L\right)$ is Knudsen number, in which $n_0=\rho_0/m$ is the number density, and $\alpha=1/\theta$. Stefanov \textit{et al}. calculated the Rayleigh number for nonrelativistic fluids as $\mbox{Ra}=2.16\left(1-r\right)/\left(\mbox{Fr}\mbox{Kn}^2\right)$ \cite{Stefanov}, which coincides with Ra under $\chi \rightarrow \infty$ in Eq. (22).\\
In this paper, we restrict ourselves to the two dimensional flow. Thus, we consider the flow on $\left(x_1,x_2\right)$-plane. Additionally, the external force ($g_i$) acts along the positive $x_1$-direction, namely, $g_2=g_3=0~\wedge~0 \le g_1$. Forms of balance equations of the momentum and energy in Eqs. (20) and (13) are equivalent to those for nonrelativistic fluids, when we set $\hat{\Pr}=\Pr$ in Eq. (20) and $\tilde{\lambda}=\lambda$ in Eq. (13). Therefore, the calculation of the critical Rayleigh number, which indicates the transition of the thermal conduction to the thermal convection, is obtained in a similar way, which is used for the calculation of the critical Rayleigh number of nonrelativistic fluids, as shown in appendix A. Actually, Stefanov \textit{et al}. \cite{Stefanov} indicated that the critical Rayleigh number, which is numerically obtained using the DSMC method, coincides with the theoretical value $1708$ in Eq. (A26) when $\mbox{Kn}=0.01$. Finally, the critical Rayleigh number of thermally relativistic fluids for three types of the boundary condition, namely, rigid (hot wall)-rigid (cold wall), slip (hot wall)-rigid (cold wall), and slip (hot wall)-slip (cold wall), are quite same as those of nonrelativistic fluids, as discussed in appendix A.\\
\textcolor{black}{In above discussion, $\theta_{\pm}$ are not defined. Then, we consider temperature jumps on two walls by calculating analytical solutions of the density and temperature under the pure conductive state in the next section.}
\section{Temperature jumps on two walls in the framework of Grad's 14 moments equation coupled with Fourier law}
\textcolor{black}{In this section, we calculate temperature jumps on two walls in the framework of Grad's 14 moments equation coupled with the Fourier law using analytical solutions of the density and temperature under the pure conductive state, where we discriminate the pure conductive state from the conductive state ($\bm{u} \neq 0$) by $\bm{u}=0$.}\\
\subsection{\textcolor{black}{Analytical solutions of density and temperature under pure conductive state}}
\textcolor{black}{Substituting $\bm{u}=0$ into Eqs. (2)-(3)}, we readily obtain following \textcolor{black}{two relations using Eqs. (4) and (5)} and $\lambda$ for hard spherical partons on the basis of Israel-Stewart theory \cite{Cercignani} \textcolor{black}{under the steady state.}
\begin{eqnarray}
&&\frac{1}{\tilde{\rho}\left(\bar{x}_1\right)}\frac{d \tilde{\rho}/\chi\left(\bar{x}_1\right)}{d\bar{x}_1}=-\frac{2}{\mbox{Fr}},\\
&&\frac{d}{d \bar{x}_1}\left[\frac{\chi\left(\bar{x}_1\right)^4 \left\{\chi(\bar{x}_1) K_2\left(\chi\left(\bar{x}_1\right)\right){}^2+5 K_3\left(\chi(\bar{x}_1)\right) K_2\left(\chi(\bar{x}_1)\right)-\chi\left(\bar{x}_1\right) K_3\left(\chi(\bar{x}_1)\right){}^2\right\}{}^2}{K_2\left(\chi(\bar{x}_1)\right){}^2 \left(5 \chi(\bar{x}_1) K_3(2 \chi(\bar{x}_1))+\left(\chi(\bar{x}_1)^2+2\right) K_2(2 \chi(\bar{x}_1))\right)}\frac{d \chi(\bar{x}_1)^{-1}}{d\bar{x}_1} \right] \nonumber \\
&&=0, \nonumber \\
\end{eqnarray}
where $\tilde{\rho}=\rho/\rho_\infty$ and $\tilde{\theta}=\theta/\theta_\infty$ are uniform in $x_2$ direction, in which $\rho_\infty$ and $\theta_\infty$ are representative values of the density and temperature. As a result, $\tilde{\rho}$ and $\tilde{\theta}$ are functions of $\bar{x}_1$.\\
Calculations of solutions of $\rho\left(\bar{x}_1\right)$ and $\chi\left(\bar{x}_1\right)$ in Eqs. (23) and (24) are markedly difficult. Thus, we calculate solutions of $\rho\left(\bar{x}_1\right)$ and $\chi\left(\bar{x}_1\right)$ under the thermally relativistic limit ($\chi \rightarrow 0$) and nonrelativistic limit ($\chi \rightarrow \infty$) in Eqs. (23) and (24).\\
From Eqs. (23) and (24), we obtain solutions of $\tilde{\rho}\left(\bar{x}_1\right)$ and $\tilde{\theta}\left(\bar{x}_1\right)$ under the thermally relativistic limit ($\chi \rightarrow 0$) as
\begin{eqnarray}
\tilde{\rho}\left(\bar{x}_1\right)&=&\frac{1}{\mathscr{C}} \breve{\rho}(\bar{x}_1),~~~\breve{\rho}(\bar{x}_1)=\left[-2c\left(a \bar{x}_1+b\right)\right]^{-1+\frac{1}{ac}},\\
\tilde{\theta}\left(\bar{x}_1\right)&=& a \bar{x}_1+b,
\end{eqnarray}
where $a=\tilde{\theta}(1)-\tilde{\theta}(0)$, $b=\tilde{\theta}(0)$, $c=-\mbox{Fr}/2$, and $\mathscr{C}=\int_0^1 \breve{\rho}\left(\bar{x}_1\right) d\bar{x}_1=(-2)^{\frac{1}{a c}}\left[(b c)^{\frac{1}{a c}}-(ca+cb)^{\frac{1}{a c}}\right]/2$. $\tilde{\theta}(0)(=\theta_{-})$ and $\tilde{\theta}(1)(=\theta_+)$ must be determined by considering the temperature jump on the wall.\\
From Eqs. (23) and (24), we obtain solutions of $\tilde{\rho}\left(\bar{x}_1\right)$ and $\tilde{\theta}\left(\bar{x}_1\right)$ under the nonrelativistic limit ($\chi \rightarrow \infty$) as
\begin{eqnarray}
\tilde{\rho}\left(\bar{x}_1\right)&=&\frac{1}{\mathscr{C}^\prime} \Check{\rho}\left(\bar{x}_1\right),~~~\Check{\rho}\left(\bar{x}_1\right)=\exp \left[-\frac{2}{3}\ln\left|a^\prime\bar{x}_1+b^\prime\right|+\frac{3 c^\prime}{a^\prime} \left(a^\prime\bar{x}_1+b^\prime\right)^{\frac{1}{3}}\right],\\
\tilde{\theta}\left(\bar{x}_1\right)&=&\left(a^\prime\bar{x}_1+b^\prime\right)^{\frac{2}{3}},
\end{eqnarray}
where $a^\prime=\tilde{\theta}(1)^{\frac{2}{3}}-\tilde{\theta}(0)^{\frac{2}{3}}$, $b^\prime=\tilde{\theta}(0)^{\frac{2}{3}}$, $c^\prime=-2/\mbox{Fr}$, and $\mathscr{C}^\prime=\int_0^1 \Check{\rho}\left(\bar{x}_1\right) d \bar{x}_1$.
\subsection{\textcolor{black}{Temperature jumps on two walls in the framework of Grad's 14 moments equation coupled with Fourier law}}
The temperature jump on the wall is explained in the framework of Grad's 13 moments under the nonrelativistic limit with the good accuracy, when $\mbox{Kn} \le 0.01$ \cite{Struchtrup}. \textcolor{black}{Here, we consider temperature jumps on two walls in the framework of Grad's 14 moments equation coupled with Fourier law using the analytical solutions of the density and temperature in Eqs. (25)-(28).}\\
\textcolor{black}{At first}, the distribution function of partons is approximated by \textcolor{black}{Grad's 14 moments equation ($f_{14}$)} on the basis of Israel-Stewart theory, \textcolor{black}{which is defined by \cite{Cercignani}}
\begin{eqnarray}
f_{14}&=&f^{(0)}\Biggl\{1+\frac{\Pi}{p}\frac{1-5 G\chi-\chi^2+G^2\chi^2}{20 G+3\chi-13 G^2\chi-2 G\chi^2+2 G^3\chi^2} \nonumber \\
&&\times \left[\frac{15 G+2\chi - 6 G^2\chi+5 G\chi^2+\chi^3-G^2\chi^3}{1-5 G\chi-\chi^2+G^2\chi^2} \right. \nonumber \\
&& +\left. \frac{3\chi}{m c^2}\frac{6 G+\chi-G^2\chi}{1-5 G\chi-\chi^2+G^2\chi^2}U_\alpha p^\alpha+\frac{\chi}{m^2 c^4} U_\alpha U_\beta p^\alpha p^\beta \right] \nonumber \\
&&+\frac{q_\alpha}{p}\frac{\chi}{\chi+5 G-G^2\chi}\left[\frac{G}{m c^2}p^\alpha-\frac{1}{m^2 c^4}U_\beta p^\alpha p^\beta \right]+\frac{\pi_{\left<\alpha\beta\right>}}{p}\frac{\chi}{2 G}\frac{1}{m^2 c^2}p^\alpha p^\beta \Biggr\}.
\end{eqnarray}
where $p^\alpha$ is the four momentum of a parton, $U^\alpha$ is the four velocity, $\Pi$ is the dynamic pressure, $\pi_{\left<\alpha \beta\right>}$ is the pressure deviator, and $f^{(0)}$ is the Maxwell-J$\ddot{\mbox{u}}$ttner function, which is defined by
\begin{eqnarray}
f^{(0)}=\frac{n}{4\pi m^2 c k\theta K_2\left(\chi\right)}e^{\frac{-U^\alpha p_\alpha}{k\theta}},
\end{eqnarray}
where $n=\rho/m$ is the number density.\\
Equation (29) is reduced to the following form using relations, $\left|\bm{u}\right|/c \ll 1$ and $\pi_{\left<ij\right>}/\left(\rho c^2\right) \ll 1$ ($i \neq j$), which is numerically confirmed from our calculation,
\begin{eqnarray}
f_{14}^{\pm}&=&f^{(0)}\Biggl[1+\frac{\tilde{\Pi}_\pm}{\tilde{n}_\pm}\frac{\chi_\pm-5G_\pm \chi^2_\pm-\chi^3_\pm+G^2_\pm\chi^3_\pm}{20 G_\pm+3\chi_\pm-13 G^2_\pm\chi_\pm-2 G_\pm\chi^2_\pm+2 G^3_\pm\chi^2_\pm} \nonumber \\
&&\times \left\{\frac{15 G_\pm+2\chi_\pm-6 G^2_\pm\chi_\pm+5 G_\pm\chi^2_\pm+\chi^3_\pm-G^2_\pm\chi^3_\pm}{1-5 G\chi_\pm-\chi^2_\pm+G^2_\pm\chi^2_\pm} \right. \nonumber \\
&& \left. + \frac{18 \chi_\pm G_\pm+3\chi^2_\pm-3 G^2_\pm\chi^2_\pm}{1-5G_\pm\chi_\pm-\chi^2_\pm+G^2_\pm\chi^2_\pm} \tilde{p}^0+\chi (\tilde{p}^0)^2 \right\} \nonumber \\
&&+\frac{\tilde{q}_1^\pm}{\tilde{n}_\pm}\frac{\chi^2_\pm}{\chi_\pm+5G_\pm-G^2_\pm\chi_\pm}\left(G_\pm \tilde{p}^1-\tilde{p}^1 \tilde{p}^0 \right)+\frac{\tilde{\pi}_{\left<ii\right>}^\pm}{\tilde{n}_\pm}\frac{\chi^2_\pm}{4 G_\pm}(\tilde{p}^i)^2\Biggr],
\end{eqnarray}
where $\chi_\pm=mc^2/\left(k\theta_\pm\right)$, $\tilde{\Pi}=\Pi/\left(n_\infty mc^2\right)$, $\tilde{\pi}_{\left<\alpha \beta\right>}=\pi_{\left<\alpha \beta\right>}/\left(n_\infty mc^2\right)$, $\tilde{p}=p/\left(n_\infty mc^2\right)$, $\tilde{q}^\alpha=q^\alpha/\left(n_\infty mc^3\right)$, $\tilde{n}=n/n_\infty$, and $\tilde{p}^\alpha=p^\alpha/(mc)$, in which the quantity with a subscript or superscript $\pm$ indicates the quantity on the cold wall ($-$) or hot wall ($+$). In Eq. (31), we used relations under the pure conductive state, namely, $U^\alpha=\left(c,0,0,0\right)$ and $q^\alpha=\left(0,q^1,0,0\right)$.\\
Distribution functions of partons, which are reflected on perfectly diffusive hot and cold walls, ($f_\pm^w$) are written as
\begin{eqnarray}
f_\pm^w=\frac{n_\pm}{4\pi m^2 c k\theta_\pm^w K_2\left(\chi_\pm^w\right)}e^{-\tilde{p}_0 \chi_\pm^w}~~~\left(p^1 \le 0\right),
\end{eqnarray}
where $\chi_\pm^w=mc^2/\left(k \theta_\pm^w\right)$ \textcolor{black}{($\theta_-^w$: temperature of the cold wall, $\theta_+^w$: temperature of the hot wall)}.\\
Continuum conditions of heat fluxes on two walls yield relations \cite{Struchtrup}
\begin{eqnarray}
q^1_{\pm}=c \int p^1 p^0 f_{14}^{\pm} \frac{d^3 p}{p^0}=c \int_{0 (<, \ge) p^1} p^1 p^0 f_{14}^{\pm} \frac{d^3 p}{p^0}+c \int_{p^1 (\le, >) 0} p^1 p^0 f_{\pm}^w \frac{d^3 p}{p^0},
\end{eqnarray}
where we used $T^{01} \simeq q^1$ under $\left|\bm{u}\right|/c \ll 1$ ($T^{\alpha \beta}$: energy-momentum tensor) \cite{Cercignani}.\\
Substituting $f_{14}^\pm$ in Eq. (31) and $f_\pm^w$ in Eq. (32) into Eq. (33), we obtain
\begin{eqnarray}
\tilde{q}^1_{\pm}&=&\pm\Biggl[\tilde{n}\frac{e^{-\chi_\pm}\left(\chi_\pm^2 +3\chi_\pm+3\right)}{\chi_\pm^3 K_2\left(\chi_\pm\right)}+\frac{\tilde{\pi}_{\left<11\right>}e^{-\chi_\pm}\left(\chi_\pm^3 + 6\chi_\pm^2 + 15\chi_\pm+15\right)}{2 \chi_\pm^3 K_3(\chi_\pm)} \nonumber \\
&& +2 e^{-\chi_\pm} \tilde{\Pi}\left\{-(\chi_\pm^5 +3\chi_\pm^4+6 \chi_\pm^3+6\chi_\pm^2) G_\pm^2 \right. \nonumber \\
&& \left. +\left(\chi_\pm^2-4\right)\left(\chi_\pm^3 +6\chi_\pm^2 + 15\chi_\pm +15\right)+\left(2\chi_\pm^4+15 \chi_\pm^3+39 \chi_\pm^2 + 39\chi_\pm\right)G_\pm\right\} \nonumber \\
&&\left\{K_2\left(\chi_\pm\right)\left[\left(4\chi_\pm^5-40\chi_\pm^3\right) G_\pm-4 \chi ^5 G_\pm^3-6 \chi_\pm^4  +26 \chi_\pm^4 G_\pm^2\right]\right\}^{-1} \nonumber \\ \nonumber \\
&& -\tilde{n} \frac{e^{-\chi_\pm^w}\left\{(\chi_\pm^w)^2 +3\chi_\pm^w+3\right\}}{(\chi_\pm^w)^3 K_2\left(\chi_\pm^w\right)} \Biggr].
\end{eqnarray}
Substituting the NSF law in Eqs. (4) and (5), and $u_1/c \simeq 0$ into Eq. (34), we obtain
\begin{eqnarray}
&&-\tilde{\lambda}_\pm \frac{d \chi_\pm^{-1}}{d \bar{x}_1}=\pm\left[\tilde{n}_\pm\frac{e^{-\chi_\pm}\left(\chi_\pm^2 +3\chi_\pm+3\right)}{\chi_\pm^3 K_2\left(\chi_\pm\right)}-\tilde{n}_\pm \frac{e^{-\chi_\pm^w}\left((\chi_\pm^w)^2 +3\chi_\pm^w+3\right)}{(\chi_\pm^w)^3 K_2\left(\chi_\pm^w\right)}\right] \nonumber \\
&&\leftrightarrow -\frac{3}{16} \mbox{Kn}\frac{\left(\chi_\pm+5 G_\pm-G_\pm^2 \chi_\pm\right)^2 \chi_\pm^4 K_2\left(\chi_\pm\right)^2}{\left(\chi_\pm^2+2\right)K_2\left(2\chi_\pm\right)+5\chi_\pm K_3\left(2\chi_\pm\right)}\frac{d \chi_\pm^{-1}}{d \bar{x}_1}= \nonumber \\
&&\pm\left[\tilde{n}_\pm\frac{e^{-\chi_\pm}\left(\chi_\pm^2 +3\chi_\pm+3\right)}{\chi_\pm^3 K_2\left(\chi_\pm\right)}-\tilde{n}_\pm \frac{e^{-\chi_\pm^w}\left\{\left(\chi_\pm^w\right)^2 +3\chi_\pm^w+3\right\}}{\left(\chi_\pm^w\right)^3 K_2\left(\chi_\pm^w\right)}\right],
\end{eqnarray}
where $\tilde{\lambda}^\pm=\lambda^\pm/\left(n_\infty c k\right)$.\\
$\chi\left(\bar{x}_1\right)$ and $\tilde{n}\left(\bar{x}_1\right)$ can be calculated under two limits, namely, $\chi \rightarrow 0$ and $\chi \rightarrow \infty$ in Eqs. (25)-(28).\\
Substituting $d \chi_\pm^{-1}/d\bar{x}_1=\chi_+^{-1}-\chi_-^{-1}$ and $n_\pm=\mbox{Fr} \chi_\pm^{-1}/\mathscr{C}$, which are obtained under the thermally relativistic limit ($\chi \rightarrow 0$) from Eq. (26), into Eq. (35), we obtain
\begin{eqnarray}
&&-\frac{3}{16} \mbox{Kn}\frac{(\chi_\pm+5 G_\pm-G_\pm^2 \chi_\pm)^2 \chi_\pm^4 K_2(\chi_\pm)^2}{(\chi_\pm^2+2)K_2(2\chi_\pm)+5\chi_\pm K_3\left(2\chi_\pm\right)}\left(\chi_+^{-1}-\chi_-^{-1}\right)=\nonumber \\
&&\pm\left[\frac{1}{\mathscr{C}}\left(\mbox{Fr}\chi_\pm^{-1}\right)^{1/ac-1}\frac{e^{-\chi_\pm}\left(\chi_\pm^2 +3\chi_\pm+3\right)}{\chi_\pm^3 K_2\left(\chi_\pm\right)}-\frac{1}{\mathscr{C}}(\mbox{Fr}\chi_\pm^{-1})^{1/ac-1} \frac{e^{-\chi_\pm^w}\left\{\left(\chi_\pm^w\right)^2 +3\chi_\pm^w+3\right\}}{\left(\chi_\pm^w\right)^3 K_2\left(\chi_\pm^w\right)}\right], \nonumber \\
\end{eqnarray}
where $\mbox{Fr}=2/\left(\tilde{g} \chi_+\right)$ ($\tilde{g}=c^2/L$).\\
Equation (36) defines $\chi_\pm$ for the pure conductive state under the thermally relativistic limit ($\chi \rightarrow 0$). Similarly, we obtain the equation for the pure conductive state from Eqs. (27), (28) and (35), which defines $\chi_\pm$ under the nonrelativistic limit ($\chi \rightarrow \infty$) as
\begin{eqnarray}
&&-\frac{1}{8} \mbox{Kn}\frac{(\chi_\pm+5 G_\pm-G_\pm^2 \chi_\pm)^2 \chi_\pm^4 K_2\left(\chi_\pm\right)^2}{\left(\chi_\pm^2+2\right)K_2\left(2\chi_\pm\right)+5\chi_\pm K_3\left(2\chi_\pm\right)}a^\prime \frac{1}{\sqrt[3]{a^\prime \Theta_\pm + b^\prime}}=\nonumber \\
&&\pm\left[\tilde{n}_\pm\frac{e^{-\chi_\pm}\left(\chi_\pm^2 +3\chi_\pm+3\right)}{\chi_\pm^3 K_2\left(\chi_\pm\right)}-\tilde{n}_\pm \frac{e^{-\chi_\pm^w}\left\{(\chi_\pm^w)^2 +3\chi_\pm^w+3\right\}}{(\chi_\pm^w)^3 K_2\left(\chi_\pm^w\right)}\right], \nonumber \\
\end{eqnarray}
where $\tilde{n}_\pm=\tilde{\rho}_\pm/m$ is obtained by Eq. (27), respectively, and $\Theta_+=1$ and $\Theta_-=0$.\\
We investigate two parameters, which evaluate temperature jumps on two walls, namely, $\Delta \theta_+=\theta_+/\theta_+^w-1=\chi_+^w/\chi_+-1$ and $\Delta \theta_-=1-\theta_-/\theta_-^w=1-\chi_-^w/\chi_-$ on two walls by setting $\chi_+^w/\chi_-^w=0.1$ ($\theta_+^w/\theta_-^w=10$) and $\sqrt{2}\mbox{Kn}=0.01$. The left frame of Fig. 2 shows $\Delta \theta_-^{14}$ ($y_1$ axis) versus $\chi_{-}^w$ ($x_1$ axis), and $\Delta \theta_+^{14}$ ($y_2$ axis) versus $\chi_{+}^w$ ($x_2$ axis) in the range of $10^{-3} \le \chi_-^w \le 10^{-1}$, which are obtained by Eq. (36), when $\tilde{g}=10^{-4}$, $100$ and $1000$, where the superscript 14 means that the value is obtained using Grad's 14 moments coupled with Fourier law. As shown in the left frame of Fig. 2, $\Delta \theta_{-}^{14}$ decreases, as $\chi_{-}^w$ ($\theta_{-}^w$) decreases (increases), when $\tilde{g}=10^{-4}$, $100$ and $1000$, whereas $d \Delta \theta_{-}^{14}/d\chi_{-}^w$ decreases, $\tilde{g}$ increases. $\Delta \theta_{+}^{14}$ slightly increases, $\chi_{-}^w$ ($\theta_{-}^w$) decreases (increases), when $\tilde{g}=10^{-4}$ and $100$, whereas $\Delta \theta_{+}^{14}$ slightly decreases, $\chi_{-}^w$ ($\theta_{-}^w$) increases (decreases), when $\tilde{g}=1000$. The right frame of Fig. 2 shows $-\Delta \theta_-^{14}$ ($y_1$ axis) versus $\chi_{-}^w$ ($x_1$ axis), and $-\Delta \theta_+^{14}$ ($y_2$ axis) versus $\chi_{+}^w$ ($x_2$ axis) in the range of $20 \le \chi_-^w \le 150$, which are obtained by Eq. (37), when $\tilde{g}=10^{-4}$, $10^{-3}$ and $0.01$. As shown in the right frame of Fig. 2, $-\Delta \theta_{-}^{14}$ decreases, as $\chi_{-}^w$ ($\theta_{-}^w$) increases (decreases), when $\tilde{g}=10^{-4}$, $10^{-3}$ and $10^{-2}$, whereas $-\Delta \theta_{-}^{14}/d\chi_{-}^w$ decreases, as $\tilde{g}$ decreases. Additionally, $-\Delta \theta_{+}^{14}$ decreases, as $\chi_{+}^w$ ($\theta_{+}^w$) increases (decreases), when $\tilde{g}=10^{-4}$ and $10^{-3}$. $-\Delta \theta_{+}^{14}$ decreases, as $\chi_{+}^w$ ($\theta_{+}^w$) increases (decreases) in the range of $2 \le \chi_{+}^w \le 6$, when $\tilde{g}=10^{-2}$, whereas $-\Delta \theta_{+}^{14}$ increases, as $\chi_{+}^w$ ($\theta_{+}^w$) increases (decreases) in the range of $6 \le \chi_{+}^w \le 15$, when $\tilde{g}=10^{-2}$.
\section{Numerical study of thermal conduction and convection of thermally relativistic fluids between two parallel walls under gravitational force}
In this section, we numerically analyze the thermal conduction and convection of thermally relativistic fluids under the gravitational force between two parallel walls by solving the relativistic Boltzmann equation on the basis of the DSMC method. \textcolor{black}{The details of the DSMC method are described in appendix B.} As the numerical condition, we consider the rectangular domain, which is stretched by $\left(\bar{x}_1,\bar{x}_2\right)=\left(\tilde{L},4\tilde{L}\right)$, as shown in Fig. 1, in which $\tilde{L}=L/L_\infty=1$. $128\times128$ grids are equally spaced in the rectangular domain. As a result, the cell size is characterized by $\Delta \bar{x}_1=1/128$ and $\Delta \bar{x}_2=1/32$. 1638400 sample particles are set in the rectangular domain. Additionally, $\sqrt{2}\mbox{Kn}=1/\left(\pi d^2 n_\infty L_\infty\right)=0.01$ indicates that flow-field is under the transition regime between the rarefied and continuum regimes, where $d$ is the diameter of a parton. Meanwhile, we must notice that the representative relaxational time $\tilde{\tau} \equiv \sqrt{2}\mbox{Kn}/\left(v_s/c\right)$ ($v_s$: speed of sound) \cite{Cercignani} increases in accordance with the increase of $\chi$, because $v_s$ decreases in accordance with the increase of $\chi$. Therefore, temperature jumps on two walls depend on both of $\chi_{\pm}^w$ and $\mbox{Kn}$. The periodic boundary condition is applied to $\bar{x}_2=0$ and $\bar{x}_2=4$, whereas partons are reflected on two perfectly diffusive walls at $\bar{x}_1=0$ and $\bar{x}_1=1$. All the numerical tests are shown in Tab. 1. As numerical tests, we consider six cases, namely, case (A) $\left(\chi_{+}^w,\chi_{-}^w\right)=\left(0.00029,0.0029\right)$, case (B) $\left(\chi_{+}^w,\chi_{-}^w\right)=\left(0.0029,0.029\right)$, case (C) $\left(\chi_{+}^w,\chi_{-}^w\right)=\left(0.01,0.1\right)$, case (D) $\left(\chi_{+}^w,\chi_{-}^w\right)=\left(0.25,2.11\right)$, case (E) $\left(\chi_{+}^w,\chi_{-}^w\right)=\left(2.11,21.1\right)$ and case (F) $\left(\chi_{+}^w,\chi_{-}^w\right)=\left(10, 100.29\right)$, as shown in Tab. 1. Each of cases includes four tests, (i)-(iv). Test (i) corresponds to the conductive state (C), test (ii) corresponds to the weakly wavy state (WW), test (iii) corresponds to the obscure formation of four vortices ($4-\mbox{V}^1$), and test (iv) corresponds to the clear formation of four vortices ($4-\mbox{V}^2$). In the next subsection, we discuss conductive states in tests (i) of cases (A)-(F).
\subsection{Numerical results of conductive state}
Table 1 shows $\Delta \theta_-$ and $\Delta \theta_+$ together with $\Delta \theta_-^{14}$ and $\Delta \theta_+^{14}$, which are calculated using Eq. (36) for cases (A)-(C) and Eq. (37) for cases (D)-(F), where we calculated $\Delta \theta_-$ and $\Delta \theta_+$ by setting $\chi_{-}=\chi\left(\Delta \bar{x}_1\right)$ and $\chi_{+}=\chi\left(1-\Delta \bar{x}_1\right)$, because we cannot calculate temperatures on two walls owing to the characteristics of the DSMC method. As shown in Tab. 1, $\Delta \theta_- \sim 0.06$ in tests (i) of cases (A)-(C), whereas $\Delta \theta_{-}^{14}$ increases, as $\chi_{-}^w$ increases in tests (i) of cases (A)-(C), as discussed in Sec. III-B. $\Delta \theta_{+}$ increases in tests (i) of cases (A)-(C), as $\chi_w^{+}$ increases, whereas $\Delta \theta_{+}^{14}$ slightly increases in tests (i) of cases (A)-(C), as $\chi_{+}^w$ decreases. The orders of $\Delta \theta_-$ and $\Delta \theta_+$ are, however, similar to those of $\Delta \theta_-^{14}$ and $\Delta \theta_+^{14}$ in tests (i) of cases (A)-(C). Additionally, $\Delta \theta_{-}$ and $\Delta \theta_{+}$ are similar to $\Delta \theta_{-}^{14}$ and $\Delta \theta_{+}^{14}$ in test (i) of case (D). Meanwhile, $\Delta \theta_{-} \ll \Delta \theta_{-}^{14}$ and $\Delta \theta_{+}^{14} \ll \Delta \theta_{+}$ are obtained in tests (i) of cases (E) and (F). Such marked differences between $\Delta \theta_{-}$ and $\Delta \theta_{-}^{14}$ or $\Delta \theta_{+}$ and $\Delta \theta_{+}^{14}$ might be explained by two reasons. One is that the representative relaxational time $\tilde{\tau}$ markedly increases in accordance with the increase of $\chi$ owing to the decrease of $v_s$, when $1 \le \chi$. As a result, the nonequilibrium effect beyond Grad's 14 moment approximation markedly increases when $\chi$ increases from $\chi=1$, even when Kn corresponds to the transition regime between the rarefied and continuum regimes. The other reason for such marked differences between $\Delta \theta_{-}$ and $\Delta \theta_{-}^{14}$ or $\Delta \theta_{+}$ and $\Delta \theta_{+}^{14}$ is that continuum conditions of $q^1$ on two walls and zero-flow condition under the pure conductive state are insufficient.\\
Figure 3 shows $\tilde{u}_1$ ($y_1$ axis) and $\tilde{\Pi}$ ($y_2$ axis) versus $1-\bar{x}_1$ in tests (i) of cases (A)-(F), where $\tilde{\bm{u}}=\bm{u}/c$. As shown in Fig. 3, the flow toward the hot wall ($1-\bar{x}_1=0$) is markedly accelerated in the vicinity of the hot wall in tests (i) of all the cases (A)-(F). Similarly, the flow toward the cold wall ($1-\bar{x}_1=1$) is also markedly accelerated in the vicinity of the cold wall in tests (i) of all the cases (A)-(F). In particular, the flow toward the hot wall is a reverse of the heat flow toward the cold wall. Such a marked acceleration in vicinities of two wall presumably yield the strongly nonequilibrium state in the vicinity wall, which does not allow us to assume that the temperature jump can be described by the continuum condition of $q^1$ on two walls under the pure conductive state on the basis of Grad's 14 moments coupled with Fourier law. $\tilde{\Pi}$ decreases in vicinities of two walls in cases (D)-(F), whereas $\tilde{\Pi}$ decreases in the vicinity of the hot wall, and markedly increases in the vicinity of the cold wall. Finally, $\tilde{\Pi} \ll 1$ is obtained in all the cases.\\
Figure 4 shows $\tilde{\pi}_{\left<11\right>}$ and $\tilde{q}^1$ versus $1-\bar{x}_1$ together with $\tilde{\pi}_{\left<11\right>}^{\mbox{\tiny{NSF}}}$ and $\tilde{q}^{1}_{\mbox{\tiny{NSF}}}$, which are calculated by substituting numerical results of $u_1$ and $\theta$ into Eqs. (4) and (5). As shown in Fig. 4, $\left|\tilde{\pi}_{\left<11\right>}^{\mbox{\tiny{NSF}}}\right| < \left|\tilde{\pi}_{\left<11\right>}\right|$ is obtained in vicinities of two walls, whereas $\tilde{\pi}_{\left<11\right>} \simeq 0$ is obtained in the regime except for vicinities of two walls. $\left|\tilde{\pi}_{\left<11\right>}\right|/\left|\tilde{\pi}_{\left<11\right>}^{\mbox{\tiny{NSF}}}\right|$ in the vicinity of the hot wall increases, as $\chi_+^w$ increases. $\left|\tilde{q}^{1}\right|<\left|\tilde{q}^{1}_{\mbox{\tiny{NSF}}}\right|$ is obtained in the vicinity of the hot wall, whereas $\left|\tilde{q}^{1}_{\mbox{\tiny{NSF}}}\right|<\left|\tilde{q}^{1}\right|$ is obtained in the vicinity of the cold wall. From above comparisons of $\tilde{\pi}_{\left<11\right>}$ and $\tilde{q}^1$ with $\tilde{\pi}_{\left<11\right>}^{\mbox{\tiny{NSF}}}$ and $\tilde{q}^{1}_{\mbox{\tiny{NSF}}}$, we confirm that strongly nonequilibrium states beyond the relativistic NSF equation are formed in vicinities of two walls in tests (i) of cases (A)-(F).\\
Here, we must confirm whether marked differences between $\Delta \theta_{-}$ and $\Delta \theta_{-}^{14}$ or $\Delta \theta_{+}$ and $\Delta \theta_{+}^{14}$ in cases (E) and (F) are dismissed by considering no permeation of the flow in normal directions to two walls ($(\tilde{u}_1)^{\pm}_w=0$) and continuum conditions of $\tilde{\pi}_{\left<11\right>}^{\pm}$ together with continuum conditions of $q^1_{\pm}$ on two walls.\\
In cases (E) and (F), we can assume that the number of partons, which satisfy the relation $\chi \tilde{p}^1\tilde{u}_1 \ll 1$, is much larger than the number of partons, which do not satisfy the relation $\chi \tilde{p}^1\tilde{u}_1 \ll 1$. As a result, we can obtain following form from Eq. (31) by setting $\tilde{\Pi} \sim 0$
\begin{eqnarray}
f_{14}^{\pm}&=&f^{(0)}_{s,\pm}\left(1+\chi_\pm \tilde{p}^1 \tilde{u}_1^\pm\right)\left[1+\frac{\tilde{q}_1^\pm}{\tilde{n}_\pm}\frac{\chi^2_\pm}{\chi_\pm+5 G_\pm-G^2_\pm\chi_\pm}\left\{G_\pm \tilde{p}^1-\tilde{p}^1 \left(\tilde{p}^0+\tilde{u}_1^\pm \tilde{p}^1 \right) \right\} \right. \nonumber \\
&&\left. +\frac{\tilde{\pi}_{\left<ii\right>}^\pm}{\tilde{n}_\pm}\frac{\chi^2_\pm}{4 G_\pm}\left(\tilde{p}^i\right)^2\right], \nonumber \\
\end{eqnarray}
where we used relations, $\tilde{U}^\alpha=U^\alpha/c \simeq \left(1,\tilde{u}_1,0,0\right)$, $\tilde{\Pi} \sim 0$ and
\begin{eqnarray}
f^{(0)}_{\pm}&=&f^{(0)}_{s,\pm} \exp\left(\chi \tilde{p}^1 \tilde{u}_1^\pm\right) \sim f^{(0)}_{s,\pm}\left(1+\chi_\pm \tilde{p}^1 \tilde{u}_1^\pm\right), \nonumber \\
f^{(0)}_{s,\pm}&=&\frac{n_\pm}{4\pi m^2 c k\theta_\pm K_2\left(\chi_\pm \right)}e^{\frac{-c p_0}{k\theta_\pm}}.
\end{eqnarray}
Zero-flow in normal directions to two walls yields relations
\begin{eqnarray}
0=c \int_{0 (<, \ge) p^1} p^1 f_{14}^{\pm} \frac{d^3 p}{p^0}+c \int_{p^1 (\le, >) 0} p^1 f_{\pm}^w \frac{d^3 p}{p^0}.
\end{eqnarray}
Continuum conditions of $\pi_{\left<11\right>}$ on two walls yield relations
\begin{eqnarray}
\pi_{\left<11\right>}^{\pm}=c \int p_1 p_1 f_{14}^{\pm} \frac{d^3 p}{p^0}=c \int_{0 (<, \ge) p^1} p_1 p_1 f_{14}^{\pm} \frac{d^3 p}{p^0}+c \int_{p^1 (\le, >) 0} p_1 p_1 f_{\pm}^w \frac{d^3 p}{p^0},
\end{eqnarray}
where we used $T_{11} \simeq \pi_{\left<11\right>}$ under $\left|\bm{u}\right|/c \ll 1$.\\
Substituting $f_{14}^\pm$ in Eq. (38) into Eq. (33), we obtain
\begin{eqnarray}
\tilde{q}^1_\pm&=&\frac{e^{-\chi_\pm} \tilde{\pi}_{\left<11\right>}^\pm \left(2 e^{\chi_\pm} \chi_\pm^4 \tilde{u}_1 K_4\left(\chi_\pm\right) \pm C_1^\pm \right)}{2 \chi_\pm^3 K_3\left(\chi_\pm\right)}
+\tilde{n}_\pm G_\pm \tilde{u}_1 \nonumber \\
&&\pm \left\{\frac{\tilde{n}_\pm e^{-\chi_\pm} C_3^\pm}{\left(\chi_\pm\right)^3 K_2\left(\chi_\pm\right)}-\frac{\tilde{n}_\pm e^{-\chi_\pm^w} C_{3,w}^\pm}{\left(\chi_\pm^w\right)^3 K_2\left(\chi_\pm^w\right)} \right\}\nonumber \\
&&+\frac{\tilde{q}^1_\pm}{B_1^\pm} \left\{3\left(\tilde{u}_1^\pm\right)^2 \chi_\pm^4  K_4(\chi_\pm) \pm e^{-\chi_\pm}\tilde{u}_1^\pm \left(4 C_2^\pm-4 \chi_\pm C_1^\pm G_\pm+2 C_1 ^\pm \right)\right\},
\end{eqnarray}
where $B_1^\pm=\chi_\pm^3 K_2(\chi_\pm)\left(\chi_\pm +5 G_\pm-\chi_\pm G_\pm^2\right)$, $C_1^\pm=\chi_\pm^3+6\chi_\pm^2+15\chi_\pm+15$, $C_2^\pm=\chi_\pm^4 +9\chi_\pm^3+39\chi_\pm^2+90\chi_\pm^4+90$ and $C_3^\pm=\chi_\pm^2+3\chi_\pm+3$.\\
Substituting $f_{14}^\pm$ in Eq. (38) into Eq. (41), we obtain
\begin{eqnarray}
\tilde{\pi}_{<11>}^\pm&=&\frac{q_\pm^1}{B_1^\pm}\left\{-3 \chi_\pm ^4 \tilde{u}_1^\pm K_2\left(\chi_\pm\right)+\chi_\pm G_\pm \left(3 \chi_\pm^3 \tilde{u}_1^\pm K_3\left(\chi_\pm\right) \pm 2 e^{-\chi_\pm} C_3^\pm \right) \right. \nonumber \\
&&\left.  \pm \left(8 \left(\tilde{u}_1^\pm\right)^2-2\right)e^{-\chi_\pm}C_1^\pm-15 \chi_\pm ^3 \tilde{u}_1^\pm K_3\left(\chi_\pm\right)\right\} \nonumber \\
&&\pm \frac{2 e^{-\chi_\pm} \tilde{n}_\pm C_3^\pm \tilde{u}_1^\pm}{\chi_\pm^3 K_2\left(\chi_\pm\right)}+\frac{\tilde{n}_\pm}{2 \chi_\pm}-\frac{\tilde{n}_\pm}{2 \chi_\pm^w}\pm \tilde{\pi}_{<11>}^\pm \left(\frac{3 e^{-\chi_\pm} C_1^\pm \tilde{u}_1^\pm}{\chi_\pm^3 K_3\left(\chi_\pm\right)}\right).
\end{eqnarray}
Substituting $f_{14}^\pm$ in Eq. (38) into Eq. (40), we obtain
\begin{eqnarray}
&&\frac{1}{2}\tilde{n}_\pm \tilde{u}_1^\pm \pm \frac{e^{-\chi_\pm } C_3^\pm \tilde{\pi}_{\left<11\right>}^\pm}{4\chi_\pm ^2 K_3\left(\chi_\pm\right)}+\frac{1}{2} \chi_\pm \tilde{\pi}_{\left<11\right>}^\pm\tilde{u}_1^\pm \pm \left\{\frac{\tilde{n}_\pm e^{-\chi_\pm } (\chi_\pm+1) }{2\chi_\pm ^2 K_2\left(\chi_\pm\right)} \right. \nonumber \\
&&\left. -\frac{\tilde{n}_\pm e^{-\chi_\pm^w} \left(\chi_\pm^w+1\right)}{2\left(\chi_\pm^w\right)^2 K_2\left(\chi_\pm^w\right)} \right\} \pm \frac{\tilde{q}^1_\pm}{{B}_1^\pm} \left\{\chi_\pm e^{-\chi_\pm} \tilde{u}_1^\pm \left(\chi_\pm+2\right)\left(C_3^\pm+3\right) \right\}+\frac{3 q^1_\pm \chi_\pm^3 \left(\tilde{u}_1^\pm\right)^2 K_3\left(\chi_\pm\right)}{2 B_1^\pm} \nonumber \\
&& \mp \frac{\tilde{q}^1_\pm}{B_1^\pm}\left(\chi_\pm^2 C_3^\pm K_3\left(\chi_\pm\right)\right)=0.
\end{eqnarray}
Finally, Fourier law, which is obtained from Eqs. (5), (27) and (28), yields the relation
\begin{eqnarray}
\tilde{q}^1_\pm=-\frac{1}{8} \mbox{Kn}\frac{\left(B_1^\pm\right)^2}{\chi_\pm^2\left(\chi_\pm^2+2\right)K_2\left(2\chi_\pm\right)+5\chi_\pm^3 K_3\left(2\chi_\pm\right)}a^\prime \frac{1}{\sqrt[3]{a^\prime \Theta_\pm + b^\prime}}.
\end{eqnarray}
From Eqs. (42)-(45), we obtain eight nonlinear equations for eight parameters ($\chi_\pm$, $\tilde{u}_1^\pm$, $\tilde{\pi}_{\left<11\right>}^\pm$ and $\tilde{q}^1_\pm$). Finally, we obtain following sets of eight parameters in cases (E) and (F)
\begin{eqnarray*}
&&\mbox{case (E)}\\
&&\chi_{-}=19.77,~\chi _+=2.15,~\tilde{q}^1_{-}=-0.004371,~\tilde{q}^1_+= -0.004222,\nonumber \\
&&\tilde{\pi}_{\left<11\right>}^-=0.007272,~\tilde{\pi}_{\left<11\right>}^+=-0.004014,~\tilde{u}_1^{-}=0.007632,~\tilde{u}_1^+=0.004955.\\
&&\mbox{case (F)}\\
&&\chi_{-}=97.15,~\chi _+=10.3,~\tilde{q}^1_{-}=-0.0004078,~\tilde{q}^1_+= -0.0004243,\nonumber \\
&&\tilde{\pi}_{\left<11\right>}^-=0.001399,~\tilde{\pi}_{\left<11\right>}^+=-0.0005426,~\tilde{u}_1^{-}=0.002209,~\tilde{u}_1^+=0.005246.
\end{eqnarray*}
From above data, we obtain $\left(\Delta \theta_-^{14},\Delta \theta_+^{14}\right)=\left(0.06712,0.0186\right)$ in case (E), and $\left(\Delta \theta_-^{14},\Delta \theta_+^{14}\right)=\left(0.03886,0.02893\right)$ in case (F). As a result, $\left(\Delta \theta_-^{14},\Delta \theta_+^{14}\right)$ in cases (E) and (F) are markedly different from that in tests (i) of cases (E) and (F). Provided that we can neglect no permeations in normal directions to two walls in Eq. (44), we obtain following sets of six parameters by setting $\left(\tilde{u}_1^-,\tilde{u}_1^+\right)=\left(-0.0121,0.0395\right)$ in case (E) and $\left(\tilde{u}_1^-,\tilde{u}_1^+\right)=\left(-0.0082,0.046\right)$ in case (F) in Eqs. (42), (43) and (45) 
\begin{eqnarray*}
&&\mbox{case (E)}\\
&&\chi_- =23.88,~\chi_+ =2.407,~\tilde{q}^1_{-}=-0.00369,~\tilde{q}^1_{+}= -0.003632,\\
&&\tilde{\pi}_{\left<11\right>}^{-}=-0.0005567,~\tilde{\pi}_{\left<11\right>}^+=-0.00448\\
&&\mbox{case (F)}\\
&&\chi_-=120.4,~\chi_+=13.28,~\tilde{q}^1_{-}=-0.0002774,~\tilde{q}^1_{+}=-0.0002873,\\
&&\tilde{\pi}_{\left<11\right>}^{-}=-0.00296,~\tilde{\pi}_{\left<11\right>}^+=-0.001212.
\end{eqnarray*}
From above data, we obtain $\left(\Delta \theta_-^{14},\Delta \theta_+^{14}\right)=\left(-0.1156,0.1232\right)$ in case (E), and $\left(\Delta \theta_-^{14},\Delta \theta_+^{14}\right)=\left(-0.1614,0.247\right)$ in case (F), which are similar to $\left(\Delta \theta_-,\Delta \theta_+\right)$ in tests (i) of cases (E) and (F), as shown in Tab. 1. Meanwhile, we must remind that artificial sets of $\left(\tilde{u}_1^-,\tilde{u}_1^+\right)$ violate no permeations in normal directions to two walls. Finally, we cannot describe temperature jumps on two walls in the framework of Grad's 14 moments coupled with Fourier law in tests (i) of cases (E) and (F). Consequently, higher order moments beyond Grad's 14 moments must be considered to describe temperature jumps on two walls owing to the increase of $\tilde{\tau}$ in tests (i) of cases (E) and (F).\\
Substituting $\theta_-$ and $\theta_+$, which are numerically obtained in tests (i) of cases (A)-(F), into Eqs. (25)-(28), $\tilde{\rho}$ versus $1-\bar{x}_1$ and $\tilde{\theta}$ versus $1-\bar{x}_1$ are plotted, as shown in Fig. 5. As shown in Fig. 5, profiles of $\tilde{\rho}$ and $\tilde{\theta}$ in tests (i) of cases (A)-(D) are similar to $\tilde{\rho}$ and $\tilde{\theta}$, which are calculated using Eqs. (25) and (26), namely, solutions of $\tilde{\rho}$ and $\tilde{\theta}$ under the thermally ultrarelativistic limit, whereas profiles of $\tilde{\rho}$ and $\tilde{\theta}$ in tests (i) of cases (E) and (F) are similar to $\tilde{\rho}$ and $\tilde{\theta}$, which are calculated using Eqs. (27) and (28), namely, solutions of $\tilde{\rho}$ and $\tilde{\theta}$ under the nonrelativistic limit. Consequently, we can conclude that the analytical solutions of the pure conductive state in Eqs. (25)-(28) are almost accurate, when we fix $\tilde{\theta}_{\pm}$ to numerical values obtained using the relativistic Boltzmann equation. Therefore, the criticism for the use of the relativistic NSF equation in Eqs. (1)-(5) owing to its acausality and instability is presumably optimistic, \textcolor{black}{when we discuss the conductive state of thermally relativistic fluids in flow-fields except for those in vicinities of two walls}. Investigating profiles of $\tilde{\theta}$ in detail, profiles of $\tilde{\theta}$, which are numerically obtained in tests (i) of cases (A)-(C), are higher than the profile of $\tilde{\theta}$, which is calculated using Eqs. (25) and (26), whereas $\tilde{\theta}$, which are numerically obtained in tests (i) of cases (A)-(C), increase more markedly toward the hot wall than $\tilde{\theta}$, which is calculated using Eqs. (25) and (26). As a result, the approximation of $\tilde{q}^1$ on the basis of Fourier law in Eq. (24) is inadequate to demonstrate the profile of $\tilde{\theta}$ around the hot wall owing to strongly nonequilibrium state around the hot wall.\\
Temperature jumps of the nonrelativistic gas on two walls can be described using the first order theory on the basis of the NSF law, as discussed by Stefanov \textit{et al} \cite{Stefanov}, when $\mbox{Kn}=0.01$. Meanwhile, our numerical results of the conductive state indicate $\left(\Delta \theta_{-},\Delta \theta_+\right)=(0.24,0.037)$ for $\mbox{Kn}=0.01$ and $\left(\Delta \theta_{-},\Delta \theta_+\right)=(0.92,0.183)$ for $\mbox{Kn}=0.1$ under the nonrelativistic limit, when $\tilde{\theta}_w^+=1.0$, $\tilde{\theta}_w^-=0.1$, $\mbox{Fr}=\infty$ and $\tilde{\tau} \sim \mbox{Kn}/\sqrt{\tilde{\theta}}$.\\
Here, we emphasize that temperature jumps on two walls under $\chi \ll 1$ cannot be described in the framework of Grad's 14 moments coupled with Fourier law even with modified transport coefficients proposed by Denicol \textit{et al}. \cite{Denicol}, which are different from those obtained on the basis of Israel-Stewart theory. 
\textcolor{black}{The transition of the conductive state to the convective state might be described in the framework of the relativistic NSF equation, when we set Kn as $\mbox{Kn} \ll 0.01$, whereas the calculation using $\mbox{Kn} \ll 0.01$ is markedly difficult because of the decrease of the time step in accordance with the decrease of Kn even with the most advanced supercomputers.}
\subsection{Numerical results of Rayleigh-B$\grave{\mbox{e}}$nard convection}
From above discussions on the conductive state, flow-fields in vicinities of two walls are under strongly nonequilibrium states, which do not allow us to describe \textcolor{black}{the transition of the conductive state to the convective state of thermally relativistic fluids in the framework of the relativistic NSF equation. Actually, temperature jumps on two walls cannot be described in the framework of Grad's 14 moments equation coupled with Fourier law, as discussed in Sec. IV-A.} Therefore, the theoretical value of the critical Rayleigh number in appendix A, which is calculated by the relativistic NSF equation in Eqs. (1)-(5), cannot characterize the transition of the conductive state to the convective state with good accuracy. We, however, use the definition of the Rayleigh number (Ra) to characterize such a transition, because we cannot introduce a new parameter other than Ra, which characterizes such a transition with better accuracy than Ra.\\
The critical Rayleigh number ($\mbox{Ra}_{\mbox{\tiny{cr}}}$), which corresponds to Ra in tests (iii), decreases, as $\chi_-^w$ ($\chi_+^w$) decreases, when $0.1 \le \chi_-^w \le 10$, whereas $\mbox{Ra}_{\mbox{\tiny{cr}}}$ decreases, as $\chi_-^w$ ($\chi_+^w$) decreases, when $2.9 \times 10^{-3} \le \chi_-^w <0.1$, as shown in Tab. 1. The critical Fr (Fr${}_{\mbox{\tiny{cr}}}$), which corresponds to Fr in tests (iii), is in the range of $30.36 \le \mbox{Fr}_{\mbox{\tiny{cr}}} \le 36.22$, whereas Fr${}_{\mbox{\tiny{cr}}} \sim 30$ under nonrelativistic limit, as obtained by Stefanov \textit{et al} \cite{Stefanov}.\\
$\Delta \theta_-$ increases, as $\tilde{g}$ increases in cases (A), (B), (D), (E) and (F), whereas $\Delta \theta_-$ under $\tilde{g}=5.4$ (test (iii)) is larger than $\Delta \theta_-$ under $\tilde{g}=5.5$ (test (iv)), as shown in Tab. 1. $\Delta \theta_+$ under the conductive state (test (i)) is larger than $\Delta \theta_+$ under the weakly wavy state (test (ii)) in cases (A)-(E), whereas $\Delta \theta_+$ under the conductive state (test (i)) is slightly smaller than $\Delta \theta_+$ under the weakly wavy state (test (ii)) in case (F), whereas $\Delta \theta_+^{14}$ under the pure conductive state increases, as $\tilde{g}$ increases, as shown in Tab. 1. Finally, $\Delta \theta_+$ increases, as $\tilde{g}$ increases from the weakly wavy state.
\section{Concluding remarks}
We investigated the thermal conduction and convection of thermally relativistic fluids under the gravitational force between two parallel walls, theoretically and numerically, \textcolor{black}{when the Lorentz contraction of thermally relativistic fluids is negligible}. Numerical results of the thermal conduction of thermally relativistic fluids under the gravitational force cannot be described in the framework of Grad's 14 moments coupled with Fourier law owing to strongly nonequilibrium states on the hot and cold walls, when the flow-field is under the transition regime between the rarefied and continuum regimes. Additionally, absolute values of temperature jumps on two walls increase, as the temperature decreases in the range of $2.9 \times 10^{-4} < \chi <120$, because the representative relaxational time increases as the temperature decreases. \textcolor{black}{Meanwhile, profiles of the density and temperature between two walls under the conductive state shows good agreements with their analytical solutions, which are obtained using the relativistic NSF equation, when we fix temperatures on two walls to those, which are numerically obtained. Therefore, we conclude that flow-fields except for those in vicinities of two walls can be described using the relativistic NSF equation owing to their weakly nonequilibrium states with good accuracy.} As a result of such strongly nonequilibrium states in vicinities of two walls beyond the relativistic NSF equation, the Rayleigh number cannot be a parameter, which characterizes the transition of the conductive state to the convective state with good accuracy, when the flow-field is under the transition regime between the rarefied and continuum regimes, whereas the Rayleigh number can be a parameter, which characterizes the transition of the conductive state to the convective state with good accuracy under the nonrelativistic limit, when the flow-field is under the transition regime between the rarefied and continuum regimes.
\begin{appendix}
\section{Critical Rayleigh number for thermally relativistic fluids}
The temperature profile is expressed with
\begin{eqnarray}
\theta\left(\bar{x}_1\right)=\theta_0+\underbrace{\bar{x}_1\left(\theta_1-\theta_0\right)+\delta \bar{\theta}^\prime\left(\bar{x}_1,\bar{x}_2\right)\left(\theta_1-\theta_0\right)}_{\delta \theta}~~~\left(0 \le \bar{x}_1 \le 1\right),
\end{eqnarray}
where $\theta_0=\theta_{-}$ and $\theta_1=\theta_{+}$.\\
To eliminate the gradient of the pressure in Eq. (20), we rewrite Eq. (20) using the vorticity. Equation (20) can be rewritten as
\begin{eqnarray}
\frac{1}{\hat{\Pr}} \left[\frac{\partial \delta \bm{\bar{u}}}{\partial \bar{t}}+\nabla\left(\frac{\delta \bar{\bm{u}} \cdot \delta \bar{\bm{u}}}{2}\right)+\delta \bar{\bm{\omega}} \times \delta \bar{\bm{u}}\right]=-\bar{\bm{\nabla}} \delta \bar{p}^\prime-\bar{\nabla}^2 \delta \bar{\bm{u}}+\mbox{Ra}\delta \bar{\theta} \bm{e},
\end{eqnarray}
where we used the relation $\delta \bar{\bm{u}} \cdot \nabla \delta \bar{\bm{u}}=\nabla \left(\frac{\delta \bar{\bm{u}} \cdot \delta \bar{\bm{u}}}{2}\right)+\delta \bar{\bm{\omega}} \times \delta \bar{\bm{u}}$.\\
Taking the curl of Eq. (A2)
\begin{eqnarray}
\frac{1}{\hat{\Pr}} \left(\frac{\partial \bar{\nabla} \times \delta \bm{\bar{u}}}{\partial \bar{t}}+\bar{\nabla} \times \delta \bar{\bm{\omega}} \times \delta \bar{\bm{u}}\right)=\bar{\nabla}^2 \left(\bar{\nabla} \times \delta \bar{\bm{u}}\right)+\mbox{Ra} \left(\bar{\nabla} \times \delta \bar{\theta} \bm{e} \right),
\end{eqnarray}
where we used the relation that the curl of a gradient is equal to zero.\\
Substituting $\delta \bar{\bm{\omega}}=\bar{\nabla} \times \delta \bar{\bm{u}}$ and $\nabla \times \delta \bar{\bm{\omega}} \times \delta \bar{\bm{u}}=\delta \bar{\bm{u}} \cdot \nabla \delta \bar{\bm{\omega}}-\delta \bar{\bm{\omega}} \cdot \nabla \delta \bar{\bm{u}}$ into Eq. (A3), we obtain
\begin{eqnarray}
\frac{1}{\hat{\Pr}} \left(\frac{\partial \delta \bar{\bm{\omega}}}{\partial \bar{t}}+\delta \bar{\bm{u}} \cdot \nabla \delta \bar{\bm{\omega}}-\delta \bar{\bm{\omega}} \cdot \nabla \delta \bar{\bm{u}}\right)=\bar{\nabla}^2 \delta \bar{\bm{\omega}}+\mbox{Ra}  \left(\bar{\nabla} \times \delta \bar{\theta} \bm{e} \right),
\end{eqnarray}
Two dimensional flow yields following relations
\begin{eqnarray}
&&\delta \bar{\omega}_3=\left(\frac{\partial \delta \bar{u}_2}{\partial \bar{x}_1}-\frac{\partial \delta \bar{u}_1}{\partial \bar{x}_2}\right)e_3,~~\delta\bar{\omega}_1=\delta\bar{\omega}_2=0,\nonumber \\
&& \delta \bar{\bm{\omega}} \cdot \nabla \delta \bar{\bm{u}}=0,\nonumber \\
&&\nabla \times \delta \bar{\theta} \bm{e}=-\frac{\partial \delta \bar{\theta}}{\partial \bar{x}_2} e_3,~~~\left(\because \bm{e}=(1,0,0)\right).
\end{eqnarray}
Equation (A4) is rewritten using Eq. (A5) as
\begin{eqnarray}
\frac{1}{\hat{\Pr}}\left(\frac{\partial \delta \bar{\omega}_3}{\partial \bar{t}}+\delta \bm{u} \cdot \bar{\bm{\nabla}} \delta \bar{\omega}_3
\right)=\bar{\nabla}^2 \delta \bar{\omega}_3-\mbox{Ra} \frac{\partial \delta \bar{\theta}}{\partial \bar{x}_2},
\end{eqnarray}
Substituting Eq. (14) and $\mathscr{U}=\tilde{\lambda}{L}$ into Eq. (13), we obtain
\begin{eqnarray}
\frac{\partial \delta \bar{\theta}}{\partial \bar{t}}+\delta \bar{\bm{u}} \cdot \bar{\nabla} \delta \bar{\theta}=\bar{\nabla}^2 \delta \bar{\theta},
\end{eqnarray}
Substituting Eq. (A1) into Eqs. (A6) and (A7), we obtain by neglecting all the nonlinear terms
\begin{eqnarray}
&&\frac{1}{\hat{\Pr}}\frac{\partial \delta \bar{\omega}_3}{\partial \bar{t}}=\bar{\nabla}^2 \delta \bar{\omega}_3-\mbox{Ra} \frac{\partial \delta \bar{\theta}^\prime}{\partial \bar{x}_2},\\
&&\frac{\partial \delta \bar{\theta}^\prime}{\partial \bar{t}}+\delta \bar{u}_1=\bar{\nabla}^2 \delta \bar{\theta}^\prime,
\end{eqnarray}
$\delta \bar{u}_1$, $\delta \bar{u}_2$ and $\delta \bar{\omega}_3$ are expressed with the stream function $\psi$ as
\begin{eqnarray}
\delta \bar{u}_1=\frac{\partial \psi}{\partial \bar{x}_2},~~~\delta \bar{u}_2=-\frac{\partial \psi}{\partial \bar{x}_1},~~~\delta \bar{\omega}_3=-\nabla^2 \psi.
\end{eqnarray}
Substituting Eq. (A10) into Eqs. (A8) and (A9), we obtain
\begin{eqnarray}
&&\frac{1}{\hat{\Pr}}\frac{\partial \nabla^2 \psi }{\partial \bar{t}}=\bar{\nabla}^2 \left(\bar{\nabla}^2 \psi\right)+\mbox{Ra} \frac{\partial \delta \bar{\theta}^\prime}{\partial \bar{x}_2},\\
&&\frac{\partial \delta \bar{\theta}^\prime}{\partial \bar{t}}+\frac{\partial \psi}{\partial \bar{x}_2}=\bar{\nabla}^2 \delta \bar{\theta}^\prime.
\end{eqnarray}
We set $\psi$ and $\delta \bar{\theta}^\prime$ as
\begin{eqnarray}
\psi&=&\hat{\psi}\left(\bar{x}_1\right)\exp\left(\sigma t-i\kappa \bar{x}_2\right),\\
\delta \bar{\theta}^\prime&=&\delta \hat{\theta}^\prime\left(\bar{x}_1\right)\exp\left(\sigma t-i\kappa \bar{x}_2\right).
\end{eqnarray}
Substituting Eqs. (A13) and (A14) into Eqs. (A11) and (A12), we obtain
\begin{eqnarray}
&&\frac{1}{\hat{\Pr}}\sigma\left(\frac{d^2}{d\bar{x}_1^2}-\kappa^2\right)\hat{\psi}=-\mbox{Ra} i\kappa \delta \hat{\theta}^\prime+\left(\frac{d^2}{d \bar{x}_1^2}-\kappa^2\right)^2\hat{\psi},\\
&&\sigma \delta \hat{\theta}^\prime-i\kappa\hat{\psi}=\left(\frac{d^2}{d \bar{x}_1^2}-\kappa^2\right)\delta \hat{\theta}^\prime.
\end{eqnarray}
$\sigma>0$ yields the instability and $\sigma<0$ yields the stability of the convection. Then, we consider the case of $\sigma=0$.\\
Substituting $\sigma=0$ into Eqs. (A15) and (A16), we obtain
\begin{eqnarray}
\kappa^2 \hat{\psi}\left(\bar{x}_1\right)=\frac{1}{\mbox{Ra}}\left(\frac{d^2}{d \bar{x}_1^2}-\kappa^2\right)^3\hat{\psi}(\bar{x}_1)
\end{eqnarray}
Substituting $\hat{\psi}\left(\bar{x}_1\right)=\exp\left(\xi \bar{x}_1\right)$ into Eq. (A17), we obtain six roots
\begin{eqnarray}
\xi_{0,1}&=&\pm i\kappa\left\{-1+\left(\frac{\kappa^2}{\mbox{Ra}}\right)^{\frac{1}{3}}\right\}^{\frac{1}{2}}, \nonumber \\
\xi_{2,3}&=&\pm i\kappa\left\{1+\left(\frac{\kappa^2}{\mbox{Ra}}\right)^{\frac{1}{3}}\exp\left(i\frac{\pi}{3}\right)\right\}^{\frac{1}{2}}, \nonumber \\
\xi_{4,5}&=&\pm i\kappa\left\{1+\left(\frac{\kappa^2}{\mbox{Ra}}\right)^{\frac{1}{3}}\exp\left(-i\frac{\pi}{3}\right)\right\}^{\frac{1}{2}},
\end{eqnarray}
From Eq. (A18), $\hat{\psi}=\sum_{i=0}^5 A_i \exp\left(\xi_i \bar{x}_1\right)$ ($i=0,1,2,3,4,5$).\\
We consider three types of the boundary condition as follows
\begin{eqnarray}
&&\mbox{Rigid boundary for two walls}\nonumber \\
&&u_2^\pm=0,~\theta(0)=\theta_{-},~\theta(1)=\theta_+.\\
&&\mbox{Slip boundary for two walls}\nonumber \\
&&\frac{d \bar{u}_2^\pm}{d\bar{x}_1}=0,~\theta(0)=\theta_{-},~\theta(1)=\theta_+.\\
&&\mbox{Slip boundary for hot wall}\nonumber \\
&&\frac{d \bar{u}_2^+}{d\bar{x}_1}=0,~\bar{u}_2^-=0,~\theta(0)=\theta_{-},~\theta(1)=\theta_+.
\end{eqnarray}
The case of the slip boundary for the hot wall and rigid boundary for the cold wall might be significant, when the density on the hot wall is rarefied by the transfer of fluids from the hot wall to cold wall.\\
From Eqs. (A15), (A16), (A19), (A20) and (A21), we can rewrite Eqs. (A19)-(A21) as
\begin{eqnarray}
&&\mbox{Equation (A19)} \rightarrow \frac{d \hat{\psi}_\pm}{d \bar{x}_1}=0,~\hat{\psi}_\pm=0,~\left(\frac{d^2}{d \bar{x}_1^2}-\kappa^2\right)^2\hat{\psi}_\pm=0,\\
&&\mbox{Equation (A20)} \rightarrow \frac{d^2 \hat{\psi}_\pm}{d \bar{x}_1^2}=0,~\hat{\psi}_\pm=0,~\left(\frac{d^2}{d \bar{x}_1^2}-\kappa^2\right)^2\hat{\psi}_\pm=0,\\
&&\mbox{Equation (A21)} \rightarrow \frac{d^2 \hat{\psi}_+}{d \bar{x}_1^2}=0,~\frac{d \hat{\psi}_{-}}{d \bar{x}_1}=0,~\hat{\psi}_\pm=0,~\left(\frac{d^2}{d \bar{x}_1^2}-\kappa^2\right)^2\hat{\psi}_\pm=0.
\end{eqnarray} 
Substituting $\hat{\psi}=\sum_{i=0}^5 A_i \exp\left(\xi_i \bar{x}_1\right)$ ($i=0,1,2,3,4,5$) into Eqs. (A22)-(A24), we obtain
\begin{eqnarray}
\mbox{M}\bm{A}=0,
\end{eqnarray}
where $\bm{A}={}^t(A_i)$ ($i=0,1,2,3,4,5$).\\
Finally, $\det \mbox{M}=0$ yields the relation between Ra and $\kappa$, as shown in Fig. 6. Finally, we obtain $\mbox{Ra}_{\mbox{\tiny{cr}}}$ for three types of the boundary condition from Fig. 6, as
\begin{eqnarray}
&&\mbox{Rigid boundary for two walls}\nonumber \\
&&\mbox{Ra}_{\mbox{\tiny{cr}}}=1708.\\
&&\mbox{Slip boundary for two walls}\nonumber \\
&&\mbox{Ra}_{\mbox{\tiny{cr}}}=657.\\
&&\mbox{Slip boundary for hot wall and rigid boundary for cold wall}\nonumber \\
&& \mbox{Ra}_{\mbox{\tiny{cr}}}=1100.
\end{eqnarray}
These values of $\mbox{Ra}_{\mbox{\tiny{cr}}}$ for three types of the boundary condition are quite same as those of nonrelativistic fluids.
\section{\textcolor{black}{Numerical method of solving relativistic Boltzmann equation}}
\textcolor{black}{The Relativistic Boltzmann equation (RBE) is written as \cite{Cercignani}}
\begin{eqnarray}
\textcolor{black}{p^{\alpha}\frac{\partial f}{\partial x^\alpha}}&\textcolor{black}{=}&\textcolor{black}{Q(f,f)} \nonumber \\
&\textcolor{black}{=}&\textcolor{black}{\int_{\mathscr{R}^3}\int_{\Omega^2} \left(f_{\ast}^\prime f^\prime -f_{\ast}f \right)F\sigma d\Omega \frac{d^3 \mathbf{p}_\ast}{p_{\ast 0}}},
\end{eqnarray}
\textcolor{black}{where $f=f\left(t,x^i,p^i\right) \ (i=1,2,3)$, and $F$ is the Lorentz invariant flux. In Eq. (B1), terms with a prime indicate conditions after collisions, $\Omega^2$ is the solid angle space and $\mathscr{R}^3$ is the momentum space stretched by $\left\{ \mathscr{R}^3 | (-\infty,-\infty,-\infty) \le (p^1,p^2,p^3) \le (\infty,\infty,\infty)\right\}$.}\\
\textcolor{black}{$x^\alpha$, $p^\alpha$ and $F$ are given by}
\begin{eqnarray}
\textcolor{black}{x^\alpha}&\textcolor{black}{=}&\textcolor{black}{\left(ct,x^1,x^2,x^3\right),} \\
\textcolor{black}{p^\alpha}&\textcolor{black}{=}&\textcolor{black}{m\gamma\left(v\right)\left(c,v^1,v^2,v^3\right)},\\
\textcolor{black}{F}&\textcolor{black}{=}&\textcolor{black}{\frac{p^0p_{\ast}^0}{c} g_{\o}\textcolor{black}{=}\frac{p^0p_{\ast}^0}{c} \sqrt{\left(\bm{v}-\bm{v}_{\ast} \right)^2-\frac{1}{c^2}\left(\bm{v}\times\bm{v}_{\ast}\right)^2} \  \  \  \  \  \left(\because g_{\o}\textcolor{black}{=}\sqrt{\left(\bm{v}-\bm{v}_{\ast} \right)^2-\frac{1}{c^2}\left(\bm{v}\times\bm{v}_{\ast}\right)^2}\right)} \nonumber \\
&\textcolor{black}{=}&\textcolor{black}{\sqrt{\left(p_{\ast}^\alpha p_{\alpha}\right)^2-m^4c^4}}.
\end{eqnarray}
\textcolor{black}{In Eq. (B3), $\gamma\left(v\right)$ is the Lorentz factor, which is defined by $\gamma\left(v\right)={1}/{\sqrt{1-v^2/c^2}}$. $c$ is the speed of light and $v^i$ ($i=1,2,3$) is the $i$th component of the particle velocity vector $\bm{v}=\left(v^1,v^2,v^3\right)$. In Eqs. (B3) and (B4), $m$ is the molecular mass. In Eq. (B4), $g_{\o}$ is M\o ller's relative velocity. In Eq. (B1), $\sigma$ is the differential cross section and $d\Omega$ is the solid angle element. In Eqs. (B1) and (B4), terms with an asterisk subscript belong to the collision partner.}\\
\textcolor{black}{Rewriting Eq. (B1) in Lorentz variant form yields}
\begin{eqnarray}
\textcolor{black}{\frac{\partial f}{\partial t}+v^i\frac{\partial f}{\partial x^i}=\int_{\mathscr{R}^3}\int_{\Omega^2}  \left(f_{\ast}^\prime f^\prime -f_{\ast}f \right)g_\phi \sigma d\Omega d^3 \mathbf{p}_{\ast}.}
\end{eqnarray}
\textcolor{black}{The direct simulation Monte Carlo (DSMC) method \cite{Bird} developed for nonrelativistic gases can be extended into relativistic gases. The Courant-Friedlichs-Lewy (CFL) condition \cite{Yee}, which is required on the left-hand side of Eq. (B1), requires that the time step $\Delta t$ approximates to zero when $p^\alpha \rightarrow \infty$, namely $v \rightarrow c$. Consequently, the left-hand side of Eq. (B1) leads to the numerical stiffness via $\Delta t \rightarrow 0$ when we consider thermally relativistic flow. Thus, Eq. (B5) is solved instead of Eq. (B1). As a numeric of the collision term in Eq. (B5), the majorant frequency scheme \cite{Ivanov} is used, whereas the past work \cite{Greiner} \cite{Bouras} used the Bird's scheme \cite{Bird}, which calculates the collision-probability for all binary collision-pairs in the numerical cell. Consequently, the computational time required by the Bird's scheme is markedly longer than that by required by the majorant frequency scheme, because the majorant frequency scheme calculates the collision-pair the maximum collision number times. In the majorant frequency scheme, the maximum collision number during $\Delta t$ is obtained for a hard-sphere particle from Eq. (5) as}
\begin{eqnarray}
\textcolor{black}{\nu_{\max}=\frac{1}{2}\left(N-1\right)n\gamma(u)\sigma_T \left(g_{\mbox{\tiny{\o}}}\right)_{\max} \Delta t=\left(N-1\right)nc \gamma(u) \sigma_T \Delta t \  \  \  (\because \left(g_{\mbox{\tiny{\o}}}\right)_{\max}=2c),}
\end{eqnarray}
\textcolor{black}{where $N$ is the number of sample particles in the cell. A collision pair is selected $\nu_{\max}$ times. The two particles selected induce a binary collision when the random number $\mathscr{W}$ ($0<\mathscr{W} \le 1$) satisfies}
\begin{eqnarray}
\textcolor{black}{\frac{g_{\mbox{\tiny{\o}}}}{(g_{\mbox{\tiny{\o}}})_{\max}}=\frac{g_{\mbox{\tiny{\o}}}}{2c} < \mathscr{W},}
\end{eqnarray}
\textcolor{black}{where $g_{\mbox{\tiny{\o}}}$ is the M$\o$ller's relative velocity for two particles selected.}\\
\textcolor{black}{Before and after the binary collision between particles 1 and 2, the total energy and total momentum of the binary particles must be conserved. The conservation of the energy $E$ and the momentum $\bm{p}=\left(p^1,p^2,p^3\right)$ before and after a binary collision is written as}
\begin{eqnarray}
\textcolor{black}{E+E_\ast}&\textcolor{black}{=}&\textcolor{black}{E^\prime+E_\ast^\prime=E_{\mbox{\tiny{tot}}}},\\
\textcolor{black}{\bm{p}+\bm{p}_\ast}&\textcolor{black}{=}&\textcolor{black}{\bm{p}^\prime+\bm{p}_\ast^\prime=\bm{p}_{\mbox{\tiny{tot}}}},
\end{eqnarray} 
\textcolor{black}{where $E=m\gamma(v) c^2$, $E_\ast=m\gamma\left(v_\ast\right)c^2$, $\bm{p}=m\gamma\left(v\right) \bm{v}$ and $\bm{p}_\ast=m\gamma\left(v_\ast\right)\bm{v}_\ast$.}\\
\textcolor{black}{$E$ and $\bm{p}$ are related as follows}
\begin{equation}
\textcolor{black}{E=\sqrt{c^2\left|\bm{p}\right|^2+m^2c^4}.}
\end{equation}
\textcolor{black}{In this paper, a binary collision is calculated using the following algorithm:}\\
\textcolor{black}{(a) Calculate the total energy ($E_{\mbox{\tiny{tot}}}$) and total momentum ($\bm{p}_{\mbox{\tiny{tot}}}=\left(p^1_{\mbox{\tiny{tot}}},p^2_{\mbox{\tiny{tot}}},p^3_{\mbox{\tiny{tot}}}\right)$) of binary particles from the left-hand sides of eqs. (B8) and (B9).}
\\
\textcolor{black}{(b) Redistribute the energy to binary particles, namely, $E^\prime$ and $E_\ast^\prime$, using a random number on the right-hand side of Eq. (B8). From eq. (B10), the norms of momenta, namely, $\left|\bm{p}^\prime\right|$ and $\left|\bm{p}_\ast^\prime\right|$, are fixed.}
\\
\textcolor{black}{(c) Decide the direction of the momentum by the law of cosines so that the total momentum is conserved in Eq. (B9) as}
\begin{eqnarray}
\textcolor{black}{\left(
\begin{array}{c}
{p^1}^\prime\\
{p^2}^\prime\\
{p^3}^\prime
\end{array}
\right)}\textcolor{black}{=\bm{M}\left(\phi,\varphi\right)\bm{N}\left(r\right)}\textcolor{black}{\left(
\begin{array}{c}
0\\
-|\bm{p}^\prime|\sin \phi_1\\
|\bm{p}^\prime|\cos \phi_1
\end{array}
\right)}, \  \  \textcolor{black}{\left(
\begin{array}{c}
{p^1_\ast}^\prime\\
{p^2_\ast}^\prime\\
{p^3_\ast}^\prime
\end{array}
\right)}\textcolor{black}{=\bm{M}\left(\phi,\varphi\right)\bm{N}\left(r\right)}\textcolor{black}{\left(
\begin{array}{c}
0\\
-|\bm{p}_\ast^\prime|\sin \phi_2\\
|\bm{p}_\ast^\prime|\cos \phi_2
\end{array}
\right),} \nonumber \\
\end{eqnarray}
\textcolor{black}{where $\phi_1$, $\phi_2$, $\bm{M}\left(\phi,\varphi\right)$ and $\bm{N}\left(r\right)$ are given by}
\begin{eqnarray}
&&\textcolor{black}{\phi_1=\arccos \left(\frac{\left|\bm{p}^\prime\right|^2+\left|\bm{p}_{\mbox{\tiny{tot}}}\right|^2-\left|\bm{p}_\ast^\prime\right|^2}{2\left|\bm{p}^\prime\right|\left|\bm{p}_{\mbox{\tiny{tot}}}\right|}\right)}, \  \  \textcolor{black}{\phi_2=\arccos \left(\frac{\left|\bm{p}_\ast^\prime\right|^2+\left|\bm{p}_{\mbox{\tiny{tot}}}\right|^2-\left|\bm{p}^\prime\right|^2}{2\left|\bm{p}_\ast^\prime\right|\left|\bm{p}_{\mbox{\tiny{tot}}}\right|}\right)}\\
&&\textcolor{black}{\bm{M}\left(\phi,\varphi\right)=\left(
\begin{array}{ccc}
\cos \phi \cos \varphi & -\sin \varphi & \sin \phi \cos \varphi\\
\cos \phi \sin \varphi & \cos \varphi & \sin \phi \sin \varphi\\
-\sin \phi & 0 & \cos \phi
\end{array}
\right)}, \  \   \textcolor{black}{\bm{N}\left(r\right)=\left(
\begin{array}{ccc}
\cos r & -\sin r& 0\\
\sin r & \cos r & 0\\
0 & 0 & 1
\end{array}
\right),}
\end{eqnarray}
\textcolor{black}{where $r$ is a random number in the range of $0 \le r \le 2\pi$, and $\phi$ and $\varphi$ are given by}
\begin{eqnarray}
&&\textcolor{black}{\phi=\arccos\left(\frac{p^3_{\mbox{\tiny{tot}}}}{\left|\bm{p}_{\mbox{\tiny{tot}}}\right|}\right),}\\
&&\mbox{\textcolor{black}{if $0 \le p^2_{\mbox{\tiny{tot}}}$}} \  \  \  \textcolor{black}{\varphi=\arccos\left(\frac{p^1_{\mbox{\tiny{tot}}}}{\left|\bm{p}_{\mbox{\tiny{tot}}}\right|\sin \phi}\right)} \nonumber\\
&&\mbox{\textcolor{black}{else}} \  \  \ \textcolor{black}{2\pi-\arccos\left(\frac{p^1_{\mbox{\tiny{tot}}}}{\left|\bm{p}_{\mbox{\tiny{tot}}}\right|\sin \phi}\right).}
\end{eqnarray}
\textcolor{black}{In our numerical analysis, 1638400 sample particles are simulated to solve the RBE in the numerical domain in Fig. 1 using 128 processors (4 nodes) of HITACHI SR 16000 through the collaborative work with Information Technology center of University of Tokyo.}
\section{Derivation of relativistic Navier-Stokes-Fourier equation in Eqs. (1)-(3)}
Balance equations of the mass, momentum density, and energy density are written in Eckart's frame as \cite{Cercignani}
\begin{eqnarray}
&&D n+n\nabla^\alpha U_\alpha=0,\\
&&\frac{n h_E}{c^2}DU^\alpha=\nabla^\alpha \left(p+\Pi\right)-\nabla_\beta \pi^{\left<\alpha\beta\right>}+\frac{1}{c^2}\Biggl(\pi^{\left<\alpha\beta\right>}DU_\beta-\Pi DU^\alpha \nonumber\\
&&-D q^\alpha-q^\alpha\nabla_\beta U^\beta-q^\beta\nabla_\beta U^\alpha-\frac{1}{c^2}U^\alpha q^\beta D U_\beta-U^\alpha \pi^{\left<\beta\gamma\right>}\nabla_\beta U_\gamma \Biggr),\\
&&n De=-\left(p+\Pi\right)+\pi^{\left<\alpha\beta\right>}\nabla_\beta U_\alpha-\nabla_\alpha q^\alpha+\frac{2}{c^2}q^\alpha DU_\alpha, 
\end{eqnarray}
where $n=\rho/m$ is the number density, $U^\alpha=\gamma(v)\left(c,v_i\right)$ ($i$=1,2,3, $\gamma\left(v\right)=1/\sqrt{1-v^2/c^2}$: Lorentz factor) is the four flow velocity, $D\equiv U^\alpha \nabla_\alpha$ is the convective time derivative, $\nabla^\alpha=\Delta^{\alpha\beta}\partial_\beta$, in which $\Delta^{\alpha\beta}=\left(\eta^{\alpha\beta}-U^\alpha U^\beta/c^2\right)\partial_\beta$, where $\eta^{\alpha\beta}=\mbox{diag}\left(1,-1,-1,-1\right)$, is the projector, $e$ is the energy density, and $h_E=mc^2 G$ is the enthalpy per a particle.\\
In this paper, we investigate thermal fluctuations under static state in the laboratory frame. Then, we assume that the product of nonlinear terms, which are expressed by products of $U_\alpha$ (or $U^\alpha$) and $\pi^{\left<\alpha\beta\right>}$, $\Pi$ or $q^\alpha$ in Eqs. (C2) and (C3), are negligible owing to $v_i \ll c$ ($i=1,2,3$).\\
Consequently, linearized balance equations of the momentum density and energy density are written from Eqs. (C2) and (C3) by neglecting nonlinear terms as \cite{Cercignani}
\begin{eqnarray}
&&\frac{n h_E}{c^2} DU^\alpha=\nabla^\alpha\left(p+\Pi\right)-\nabla_\beta \pi^{\left<\alpha\beta\right>}-\frac{1}{c^2}D q^\alpha,\\
&&n D e=-p\nabla_\beta U^\beta-\nabla_\beta q^\beta.
\end{eqnarray}
In Eq. (C4), we assume that th term $-Dq^\alpha/c^2$ is negligible, and rewrite Eq. (C4) as
\begin{eqnarray}
\frac{n h_E}{c^2} DU^\alpha=\nabla^\alpha\left(p+\Pi\right)-\nabla_\beta \pi^{\left<\alpha\beta\right>}.
\end{eqnarray}
From Eqs. (C1), (C5) and (C6), we readily obtain Eqs. (1)-(3) using relations, $De=c_v D\theta$, $h_E=mc^2 G$, and $\gamma\left(v\right)=1$.
\end{appendix}

\newpage
\begin{table}[hbtp]
\caption{Numerical results of cases (A)-(F)}
\label{table:data_type}
\begin{tabular}{l||cccccccc}
\hline
Case (A) & $\tilde{g}$ & $\Delta \theta_{-}$ & $\Delta \theta_{+}$ & $\Delta \theta_{-}^{14}$ (Eq. (36)) & $\Delta \theta_{+}^{14}$ (Eq. (36)) & Ra & Fr &  Status\\
\hline \hline
test (i) & $100$ & 0.059 & 0.01442 & 0.03352 & 0.02605 & 437.7 & 68 & C \\
test (ii) & $193.5$ &  0.06285	& 0.0137 & 0.03352 & 0.02605 &	846.1	& 35.15 & WW \\
test (iii) & $194$ & 0.0681 & 0.01485 & 0.03352 & 0.02605 & 848.6 & 35.02 & 4-V${}^1$\\
test (iv) & 195	& 0.075 & 0.02 & 0.03352 & 0.02605 & 856.2 & 34.66 & 4-V${}^2$\\
\hline
Case (B) & $\tilde{g}$ & $\Delta \theta_{-}$ & $\Delta \theta_{+}$ & $\Delta \theta_{-}^{14}$ (Eq. (36)) & $\Delta \theta_{+}^{14}$ (Eq. (36)) & Ra & Fr & Status\\
\hline \hline
test (i) &  0 & 0.064 & 0.0155 & 0.05768 & 0.02579 & 0 & $\infty$ & C\\
test (ii) & 19.15 & 0.068 & 0.01222 & 0.05766 & 0.0258  & 835.7 & 35.57 & WW\\
test (iii) & 19.25 & 0.07 & 0.01271 & 0.05766 & 0.0258 & 840.3 & 35.38 & 4-V${}^1$\\
test (iv) &19.5 & 0.076 & 0.017 & 0.05766 & 0.0258 &  853.8 & 34.76	& 4-V${}^2$\\
\hline
Case (C) & $\tilde{g}$ & $\Delta \theta_{-}$ & $\Delta \theta_{+}$ & $\Delta \theta_{-}^{14}$ (Eq. (36)) & $\Delta \theta_{+}^{14}$ (Eq. (36)) & Ra & Fr & Status\\
\hline \hline
test (i) & $1$ & 0.06 & 0.022 & 0.1264 & 0.02508 & 151.9 & 195.6 & C \\
test (ii) & 5.35 & 0.0665 & 0.02031 & 0.1264 & 0.0251 & 811.1 & 36.62 & WW \\
test (iii) & 5.4 &  0.07143 & 0.02186 & 0.1264  & 0.0251 & 819.3 & 36.22 & 4-V${}^1$\\
test (iv) & 5.5 &  0.069 & 0.0222 & 0.1264 & 0.0251 & 835 & 35.6 & 4-V${}^2$\\
\hline
Case (D) & $\tilde{g}$ & $\Delta \theta_{-}$ & $\Delta \theta_{+}$ & $\Delta \theta_{-}^{14}$ (Eq. (37)) & $\Delta \theta_{+}^{14}$ (Eq. (37)) & Ra & Fr & Status\\
\hline \hline
test (i) & 0.1 & -0.01927 & 0.0336 & -0.02285 & 0.02309 & 381 & 77.31& C\\
test (ii) & 0.235 & -0.01674 & 0.033 & -0.02244 & 0.0235 & 894.5 & 32.92 & WW \\
test (iii) & 0.24 & -0.01252 & 0.0337 & -0.02243 & 0.02351 &913.6 & 32.21 & 4-V${}^1$\\
test (iv) & 0.25 & 0.00014 & 0.036614 & -0.0224 & 0.02355 & 952.4 & 30.83 & 4-V${}^2$\\
\hline
Case (E) & $\tilde{g}$ & $\Delta \theta_{-}$ & $\Delta \theta_{+}$ & $\Delta \theta_{-}^{14}$ (Eq. (37)) & $\Delta \theta_{+}^{14}$ (Eq. (37)) & Ra & Fr & Status\\
\hline \hline
test (i) & 0.01 & -0.12 & 0.121 & -0.000234 & -0.00954 & 395.1 & 83.3 & C\\
test (ii) & 0.025 & -0.11 & 0.105 & -0.0002054 & -0.01106 & 969.6 & 33.92 & WW \\
test (iii) & 0.0275 & -0.11 & 0.1193 & -0.0002012 & -0.01135 & 1083 & 30.36 & 4-V${}^1$ \\
test (iv) &0.03 & -0.076 & 0.1248 & -0.000197 & -0.01164 & 1184 & 27.65 & 4-V${}^2$\\
\hline
Case (F) & $\tilde{g}$ & $\Delta \theta_{-}$ & $\Delta \theta_{+}$ & $\Delta \theta_{-}^{14}$ (Eq. (37)) & $\Delta \theta_{+}^{14}$ (Eq. (37)) & Ra & Fr & Status\\
\hline \hline
test (i)  & 0.0025 & -0.16 & 0.2494 & -6.878$\times 10^{-6}$ & -0.0006361 & 608.24 & 60.05 & C\\
test (ii) & 0.0045 & -0.159 & 0.25 & -4.984$\times 10^{-6}$ & -0.001073 & 1095 & 33.33 & WW \\
test (iii) & 0.00475 & -0.15 & 0.2513 & -4.805$\times 10^{-6}$ & -0.001149 & 1290 & 31.52 & 4-V${}^1$ \\
test (iv) & 0.005 & -0.14 & 0.2527 & -4.636$\times 10^{-6}$ & -0.001232 & 1218 & 29.9 & 4-V${}^2$ \\
\hline
\end{tabular}
\end{table}
\newpage
\hspace{-1.5em}\textbf{Figure captions}\\
Figure 1: Schematic of flow-field.\\
Figure 2: $\Delta \theta^{14}_-$ versus $\chi_{-}^w$ and $\Delta \theta^{14}_+$ versus $\chi_+^w$ obtained using Eq. (36), when $\tilde{g}=10^{-4}$, $100$ and $1000$ (left frame). $\Delta \theta^{14}_-$ versus $\chi_{-}^w$ and $\Delta \theta^{14}_+$ versus $\chi_+^w$ obtained using Eq. (37), when $\tilde{g}=10^{-4}$, $10^{-3}$ and $10^{-2}$ (right frame).\\
Figure 3: $\tilde{u}_1$ versus $1-\bar{x}_1$ in $y_1$ axis and $\tilde{\Pi}$ in $y_2$ axis versus $1-\bar{x}_1$ in tests (i) of cases (A)-(F).\\
Figure 4: $\tilde{q}^1$, $\tilde{q}^1_{\mbox{\tiny{NSF}}}$, $\tilde{\pi}_{\left<11\right>}$,  and $\tilde{\pi}_{\left<11\right>}^{\mbox{\tiny{NSF}}}$ versus $1-\bar{x}_1$ in tests (i) of cases (A)-(F).\\
Figure 5: $\tilde{\theta}$ versus $1-\bar{x}_1$ in $y_1$ axis and $\tilde{\rho}$ versus $1-\bar{x}_1$ in $y_2$ axis, where analytical solutions of $\tilde{\theta}$ and $\tilde{\rho}$ in Eqs. (25)-(28) are calculated using $\theta_{-}$ and $\theta_{+}$, which are numerically obtained.\\
Figure 6: Ra versus $\kappa$ for three types of the boundary condition, where shaded areas correspond to instable regimes.
\end{document}